# Inter-Nanoparticle Repulsion Effect on Self-Organization in Polymer Nanocomposites: A Langevin Dynamics Simulation Study

**Paawan Chandrakanth[1], Nikos Ch. Karayiannis[2], Subhalaxmi Das[1] and Supriya Roy [1,a]**

[1] School of Applied Sciences, Kalinga Institute of Industrial Technology, Bhubaneswar, Odisha, India

[2] Institute for Optoelectronic Systems and Microtechnology (ISOM) and Escuela Técnica Superior de Ingenieros Industriales (ETSII), Universidad Politécnica de Madrid (UPM), José Gutiérrez Abascal 2, E-28006 Madrid, Spain

[a] Author to whom correspondence should be addressed: supriyophy163@gmail.com

Demixing through liquid-liquid phase separation is relevant to nanoparticle (NP) aggregation in polymer nanocomposites (PNCs), biomolecular compartmentalization, and liquids embedded in elastic networks. Mixing-demixing phase behavior depends on factors, such as particle size distribution, system dimensionality, intermolecular interactions, and relative concentrations. Here, using Langevin Dynamics simulations, we systematically investigate the effect of inter-nanoparticle repulsion on the self-assembly of PNCs under extreme confinement. A nanoparticle assembly, in the form of a compact droplet, is immersed in a polymer matrix of semiflexible, rod-like chains. By varying the nanoparticle-nanoparticle repulsion strength, while fixing all other interactions, we identify four distinct regimes with respect to the self-assembled NP morphologies: droplet, slab, crystalline droplet, and dispersed. Cluster-based analysis reveals a progressive loss of nanoparticle droplet integrity, leading to fragmentation within the dispersed regime. Local order of the nanoparticle assembly and global order of polymer chains are gauged through refined descriptors, quantifying the degrees of crystallinity and of nematic order, respectively. Interestingly, the slab regime exhibits negligible nanoparticle crystallization yet demonstrates the strongest polymer nematic ordering. In contrast, the crystalline droplet displays extensive triangular close-packed crystallites accompanied by a significant reduction in polymer nematic order, owing to its rhomboid shape. In the dispersed regime, nanoparticle fragmentation facilitates efficient mixing and local smectic-like ordering of the polymer chains. These findings demonstrate that fine-tuning inter-NP repulsions can control self-organization and the resulting morphologies, providing insights into PNCs with tailored functionality.

## 1. INTRODUCTION

Liquid–liquid phase separation (LLPS) is relevant to a variety of systems, such as polymer nanocomposites (PNCs),[1,2] liquid mixtures within elastic polymer[3] and fibrillar[4] networks, and the compartmentalization of membrane-less organelles in cells.[5,6] In PNCs, nanoparticles (NPs) are embedded in a polymer matrix to improve its mechanical, optical, and electrical properties.[2,7] The spatial organization of the NPs plays an important role in this functionality: a homogeneous dispersion maximizes the polymer-NP interfacial area, whereas phase separation often degrades performance[8,9]. Achieving and maintaining NP dispersion has therefore been a central challenge in PNC engineering and has attracted many experimental, theoretical, and simulation efforts[10–20]. In living cells, on the other hand, membrane-less organelles exhibit liquid-like properties and often coalesce into biomolecular condensates through LLPS, which helps the organization of cellular components[6,21]. For example, P-granule droplets[22] condense in the posterior of the *Caenorhabditis elegans* embryo. Stress granules, complexes of proteins and mRNA, grow in the cytoplasm under cellular stress, a process facilitated by microtubules[23].

First proposed by Asakura and Oosawa[24,25] and later independently formulated by Vrij[26], the Asakura-Oosawa-Vrij model demonstrates that the overlap of the excluded volumes of nanoparticles (NPs) generates an osmotic depletion attraction that can drive NP phase separation in polymer-NP mixtures. For hard-sphere NPs, this phase separation is driven by the entropy gained by the polymers. Subsequent theoretical[27–29] and experimental works[30,31] on colloid-polymer mixtures observed fluid-solid coexistence and three-phase coexistence of colloidal gas, liquid, and crystal. The phase diagram is largely governed by the size ratio of the polymer radius of gyration to the colloid diameter[30]. For rod-like colloids such as tobacco mosaic virus, the depletion attraction promotes nematic ordering[32]. Added spheres can either stabilize or disrupt the nematic phase and induce rich rod-sphere demixing, depending on the sphere-to-rod size ratio and particle concentration[33–38]. Recent experiments[39] on fd-virus and polystyrene sphere mixtures, combined with free volume theory and simulation[40], have mapped how rod length, rod diameter, and sphere radius govern these phase boundaries.

When the NPs are not treated as hard spheres, the phase behavior is no longer set by entropy alone. The interaction energy between the different species, in all possible combinations, also contributes significantly to the free energy, so whether the NPs disperse, aggregate, or crystallize depends on the balance between the polymer-mediated depletion attraction and the direct energetic interactions[41–44]. Past coarse-grained simulations have revealed the effect of polymer-filler interactions on filler dispersion[13]. Another simulation study of adsorbing polymers[45] found that NP dispersion and aggregation depend on polymer concentration and polymer-NP attraction strength, which are linearly related at the phase boundary between the two states. Molecular dynamics simulations using coarse-grained models explored the aggregation and dispersion of cubic and tetrahedral NPs as a function of polymer-NP interaction strength and NP volume fraction[46].

The presence of confinement adds another dimension to the complexity in the study of self-organization in PNCs [47–49]. Simulations[50,51] of polymer-NP mixtures between wall have examined the crossover from surface segregation to bulk aggregation with volume fraction, particle size and shape, controlled by particle-wall and particle-particle depletion attractions. Experimentally, polymer nanocomposite thin films resemble quasi-2D polymer-NP mixtures, and coarsening of ligand-stabilized gold NPs has been observed in thin poly (methyl methacrylate) films[52,53]. Simulations of semiflexible polymers under strong slit-like confinement[53], smaller than the persistence length, revealed second-order isotropic-nematic[54] and nematic-smectic[55] transitions compared to the first order one in bulk solutions[56].

To the best of our knowledge, the phase and structural behavior of a polymer-NP mixture under strong confinement as function of the inter-NP repulsion has not been explored, apart from the two limiting cases in our earlier study[57]. In that work on strongly confined semiflexible polymer-NP mixtures, we studied hard NPs (HNPs) that repel each other through the Weeks-Chandler-Andersen (WCA) potential, and ideal NPs (INPs) that do not interact with one another. The two limits lead to different behaviors. HNPs remain dispersed at low and moderate volume fractions and phase separate into crystalline domains at higher volume fractions. In parallel, the nematic ordering is disrupted and the isotropic–nematic transition is shifted to larger monomer volume fractions. INPs, in contrast, phase separate at moderate and high-volume fractions, form oval shaped droplets and rectangular slabs, and facilitate the isotropic-nematic transition by lowering the critical concentration. Since overlapping INPs costs no energy, the ideal limit is an extreme case of soft particles, whereas HNPs represent strongly repulsive particles, such as inorganic NP fillers, that restrict aggregation. This raises a natural question of how the system evolves between the ideal and hard limits.

Between these two limits lies a variety of soft particles, such as soft colloids, star polymers, microgels, and emulsions, which allow substantial overlap depending on their degree of softness[58,59]. Coiled polymers are regarded as the softest and emulsions as the least soft among them. Previous models of ultrasoft particles[60,61], such as the penetrable sphere[62], Fermi-distribution[63], and Gaussian core[64], generalized exponential of index $n$[65], use purely repulsive potentials that remain finite even at complete overlap. In the present study, we instead tune the NP-NP repulsion by varying the strength of the WCA potential from zero (ideal limit) to unity (hard limit). Since the repulsion strength controls the extent of NP overlap, this allows us to monitor the phase behavior of NPs ranging from soft, organic-like additives that tend to coalesce to inorganic hard fillers that resist aggregation. It also enables identification of the phase boundary between coalesced, crystalline and dispersed states of NPs in the semiflexible polymer matrix, as well as the effect of these states on the polymer nematic phases. Towards this, we employ GPU-accelerated Langevin dynamics (LD), together with refined metrics to quantify cluster formation.

We then gauge both the local structures at the NP level and the nematic order exhibited by the polymer chains. In all cases, initial configuration corresponds to a droplet-like structure of NPs inserted into the polymer matrix of semiflexible, rod-like chains under extreme confinement in one dimension. The model studied here is not intended to represent a specific biological or experimental system. Rather, it explores, at a fundamental level, how the competition among depletion attraction, inter-NP repulsion, and polymer alignment leads to rich phase behavior and self-organization into different morphologies.

The manuscript is organized as follows: Section 2 describes the model and the simulated systems; Section 3 presents the main results; and Section 4 summarizes the conclusions and our current research efforts.

## 2. MATERIALS AND METHODS

Systems comprising a mixture of spherical NPs and semiflexible chains under strong confinement are simulated using GPU-accelerated LD simulations in the canonical NVT ensemble (constant number of sites, *N*, volume, *V*, and temperature, *T*). The Langevin equation includes stochastic and frictional forces in addition to the conservative forces to represent the effects of an implicit solvent:

$$m_i \frac{d^2\boldsymbol{r_i}}{dt^2} = -\vec{\nabla}_i U - \gamma_i \frac{d\boldsymbol{r_i}}{dt} + \boldsymbol{R_i}(t) \tag{1}$$

where $m_i$ is the mass of particle $i$, $\gamma_i$ is the friction coefficient, $\boldsymbol{r_i}$ is the coordinate vector and $\boldsymbol{R_i}$ is the stochastic force arising from the thermal fluctuation of the implicit solvent. $U$ and $t$ correspond to the energy function and time, respectively. The stochastic force satisfies the fluctuation-dissipation relation and the integration employs the Velocity-Verlet algorithm, as described in the work by Paterlini and Ferguson[66].

As the system is a composite made of two distinct constituents, throughout the manuscript we will use the indices "*m*" and "*p*" to denote monomer bead and nanoparticle (NP), respectively. Each polymer is composed of $N_m = 25$ monomer beads of uniform diameter $\sigma_m$. The monomer diameter $\sigma_m$ is used as the fundamental unit of length, and all lengths and distances reported in the continuation are expressed in units of $\sigma_m$ ($\sigma_m$ = 1). The diameter of the NPs is uniformly set to $\sigma_p = 2.5$. The mass of a monomer bead $m_m$ is set to 1 and used as the unit of mass, while the thermal energy $k_BT$, where $k_B$ is the Boltzmann constant, is set as the unit of energy. $\gamma$ for monomer bead is set to 1; consequently, the monomer bead diffusion time is $\tau_D = \gamma {\sigma_m}^2/k_BT = 1$. Without loss of generality, and considering that the focus of the present study lies on the structural changes rather than dynamics, the NPs are assigned the same mass and friction coefficient as the monomers. The integration timestep is $\Delta t = 0.02\tau_D$.

The potential energy of the system, $U$, is a sum of four contributions, namely $U_{spring}$, $U_{bend}$, $U_W$ and $U_{EV}$ as:

$$U = U_{spring} + U_{bend} + U_w + U_{EV} \tag{2}$$

Consecutive monomer beads within a polymer are bonded through a harmonic spring potential of the form:

$$\frac{U_{spring}}{k_BT} = \frac{K_V}{2}(r - \sigma_m)^2 \tag{3}$$

With $K_V = 400$, where $r$ is the distance between the bonded monomer beads. To account for chain stiffness, a bending potential of the form

$$\frac{U_{bend}}{k_BT} = K_B(1 - \cos\theta) \tag{4}$$

is introduced, where $\theta$ is the bending angle between two successive monomer-monomer bonds within the same polymer, with $K_B = 20$ representing the bending rigidity. No torsional potential is applied.

A parallelopiped simulation box is chosen with dimensions $L_x = L_y = 100$ and $L_z = 4$. Periodic boundary conditions are applied along the $x$ and $y$ dimensions while confinement is applied on the short $z$ dimension, corresponding practically to an extremely confined thin-film PNC. Confinement is realized through a repulsive potential, $U_{wi}$, between each flat wall and particle $i$, which represents either a monomer bead or an NP:

$$\frac{U_{W_i}}{k_B T} = \begin{cases} k_w \sigma_i \left(1 - \frac{r_{wi}}{\sigma_i}\right)^3, & r_{wi} < \sigma_i \\ 0, & r_{wi} \geq \sigma_i \end{cases} \tag{5}$$

where $k_w$ is the constant of the repulsive energy and $r_{wi}$ is the minimum distance between the wall and the center of the particle $i$. In all simulations we set $k_w = 50$. As seen in Eq. (4), compared to our previous work[57], the potential coefficient is multiplied by the diameter of the species (monomer beads or NPs) to ensure a stricter confinement for the larger particles (i.e. the NPs).

All non-bonded interactions are modelled using the repulsive Weeks-Chandler-Andersen (WCA) variation of the shifted and truncated Lennard-Jones type potential as implemented[67] in the study of polymer-colloidal particle mixture by Doxastakis *et al.*[68]:

$$\frac{U_{EV,ij}}{k_B T} = \begin{cases} 4\varepsilon_{ij}\left\{\left[\frac{\sigma_{ij}}{(r_{ij} - D_{ij})}\right]^{12} - \left[\frac{\sigma_{ij}}{(r_{ij} - D_{ij})}\right]^{6}\right\} + \varepsilon_{ij}, & r_{ij} - D_{ij} \leq 2^{1/6}\sigma_{ij} \\ 0, & r_{ij} - D_{ij} > 2^{1/6}\sigma_{ij} \end{cases} \tag{6}$$

where $U_{EV,ij}$ is the non-bonded potential energy of the pair of particles $i$ and $j$, whose centers are separated by a distance $r_{ij}$. $\varepsilon_{ij}$ and $\sigma_{ij}$ are the repulsive intensity and collision distance, respectively. In the equation above $D_{ij} = |\sigma_i - \sigma_j|/2$ and $\sigma_{ij} = (\sigma_i + \sigma_j)/2$. In this system, three different types of pair combinations for particles $i$ and $j$ are possible: $mm, pp,$ and $pm$. It should be noted that non-bonded interactions are applied to all $mm$ pairs irrespective of having bonded interactions in the form of a bond spring (Eq. 3) or bending potential (Eq.4).

In all reported simulations we set $\varepsilon_{mm} = \varepsilon_{pm} = 1$, while the corresponding one for the NP-NP pairs, $\varepsilon_{pp}$ is systematically varied from 0 to 1 with a step size varying from 0.0125 to 0.1, depending on the region of the phase diagram explored. Very small step sizes are chosen in the vicinity of structural changes to identify with precision the transition thresholds. For $\varepsilon_{pp} = 0$ (Ideal NPs), NPs do not mutually interact, whereas, in the other limit corresponding to $\varepsilon_{pp} = 1$ (Hard NPs), NPs strongly repel each other[57]. In the Hard NP limit it further holds that $\varepsilon_{mm} = \varepsilon_{pm} = \varepsilon_{pp} = 1$, so that all pairs share the same repulsion parameter, independent of type.

In a system containing $M$ polymer chains (each with $N_m$ monomers of diameter $\sigma_m$) and $N_p$ nanoparticles (diameter $\sigma_p$), the volume fractions are computed using the hard-sphere (HS) approximation as $\phi_m$ and $\phi_p$, for monomers and NPs, respectively, according to:

$$\phi_m = \frac{\pi \sigma_m{}^3 M N_m}{6V}, \quad \phi_p = \frac{\pi \sigma_p^3 N_p}{6V} \tag{7}$$

Alternatively, given that hard confinement does not permit particles lying closer than their radius to the nearest wall an effective packing density can be introduced[56,69] to account for spatial constraints as:

$$\phi_m^{eff} = \frac{\pi \sigma_m^3 M N_m}{6 L_x L_y (L_z - \sigma_m)}, \quad \phi_p^{eff} = \frac{\pi \sigma_p^3 N_p}{6 L_x L_y (L_z - \sigma_p)} \tag{8}$$

In a previous simulation study, Roy *et al*[57] reported that the addition of Ideal NPs ($\varepsilon_{pp} = 0$) into an already nematic polymer matrix led to phase separation and the formation of an oval-shaped NP droplet. Under the same confinement $L_z = 4$, for a pure semiflexible polymer system with bending rigidity $K_B = 20$, Luzhbin *et al.*[54] reported a critical monomer volume fraction of $\phi^{I-N} = 0.096$ for the isotropic-to-nematic (I-N) transition. Since our objective is to investigate the behavior of a droplet-like NP aggregate immersed in a polymer matric under varying inter-NP repulsion, we use $M = 500$ polymers, corresponding to $\phi_m = 0.164$ ($\phi_m^{eff} = 0.218$), which is sufficiently above $\phi^{I-N}$, ensuring the establishment of nematic order. The number of nanoparticles is $N_p = 800$, corresponding to $\phi_p = 0.164$ $(\phi_p^{eff} = 0.436)$. In all simulations we keep the number of polymer chains, $M$, the total number of monomer beads, $MN_m$, and the population of nanoparticles, $N_p$, the same. For the $\varepsilon_{pp} = 0$ system, the NPs and the polymers are inserted at random positions, and the NPs spontaneously phase separate and coalesce into a single, compact NP droplet within $3.3 \times 10^7$ timesteps. The configuration of the Ideal NP after $7 \times 10^7$ timesteps is then used as the initial configuration for all simulations. The LD simulations are extended for $4 \times 10^8$ timesteps, ensuring thorough equilibration. System equilibration is verified by tracking the temporal evolution of the total energy and of the pair radial distribution functions, as well as the stability of the resulting morphologies for NPs and polymers. This is further supported by surveying the evolution of the metrics used to gauge local and global order, to be presented below. Most systems reach equilibration at around $10^8$ steps, while selected ones require longer times ($2 \times 10^8$). System configurations are saved at an interval of either $1 \times 10^6$ or $2 \times 10^6$ timesteps, depending on the system (400 or 200 configurations per run). The final part of the simulation trajectory, composed of 100 frames, is used for the statistical analysis. Each simulation runs on a single GPU (Nvidia A100) and for the given system size a simulation of $4 \times 10^8$ timesteps requires roughly one day of wall-clock time.

We should note here that systems with smaller $\varepsilon_{pp}$ values exhibit weaker repulsive interactions between NPs, leading to partial overlaps between them, a trend which weakens as $\varepsilon_{pp}$ increases. This overlap effectively reduces the system's volume fraction compared to that estimated using the hard-sphere approximation. To accurately determine the effective volume fraction, we employ a Voronoi tessellation analysis, which is described in detail in the Results and Discussions section.

First, we visualize the computer-generated system configurations which allow for an inspection of the established morphologies. Second, we calculate the pair radial distribution function (RDF) in all possible combinations ("$m-m$", "$m-p$" "and "$p-p$"). Voronoi tessellation follows, which identifies the Voronoi volume around each NP and monomer bead as the inverse of the local packing density. At the level of NPs, first we perform a cluster analysis based on proximity criterion, as implemented in the Simu-D software[70–72], to quantify the level of aggregation/dispersion. Then, we analyze the local structure and the corresponding degree of ordering through the characteristic crystallographic element (CCE) norm[73,74]. Global order, at the level of polymer chains, is quantified through the nematic order parameter, as in our past studies[55,57].

## 3. RESULTS AND DISCUSSION

### *3.1. Self-Organization: System configurations*

Figure 1 shows the resulting system configurations at the end of the Langevin simulations as repulsion intensity between NPs increases progressively from 0 (Ideal NP) to 1 (Hard NP). Based on a combination of visual inspection and quantification through diverse metrics, as presented below, four distinct phases can be identified with respect to the NP morphologies. The NP regimes, with the corresponding nanostructures and the threshold repulsion values, are:

i. overlapping droplet ($0 \leq \varepsilon_{pp} \leq 0.025$); see panels (a)-(c) in Figure 1;
ii. slab $(0.025 < \varepsilon_{pp} \leq 0.1)$; see panels (d)-(g) in Figure 1;
iii. crystalline droplet $(0.1 < \varepsilon_{pp} \leq 0.225)$; see panels (h)-(l) in Figure 1;
iv. dispersion $(0.225 < \varepsilon_{pp} \leq 1)$; see panels (m)-(t) in Figure 1.

At weak repulsion, (regime i, $0 \leq \varepsilon_{pp} \leq 0.025$), an interface is formed between NPs and polymers, with the former remaining aggregated into a single dense droplet. Since $\varepsilon_{pm} = 1$ in all cases studied here, this NP-polymer phase is driven by very weak repulsive, or even ghost (for $\varepsilon_{pp} = 0$) interactions between NPs compared to the stronger NP-monomer ones. This results in an energetically favorable configuration with extensive overlaps between NPs in the droplet domain, due to the greatly reduced energetic penalty. Therefore, phase separation is observed into polymer-rich and NP-rich regions. As expected, the self-assembled NP droplet obtains its minimum size and thus its densest state for $\varepsilon_{pp} = 0$. Once the repulsion is activated between NPs ($\varepsilon_{pp} > 0$), NP overlapping becomes energetically more expensive, leading to droplet expansion and a more elliptical shape, as is visually evident in the snapshots in Figure 1(a)-(c). The change in shape is attributed to the increasing energy cost of wrapping a larger NP domain, which requires, in addition to the non-bonded interactions, more bending energetic compensation as the extended polymer conformations lie parallel to the contour of the droplet.

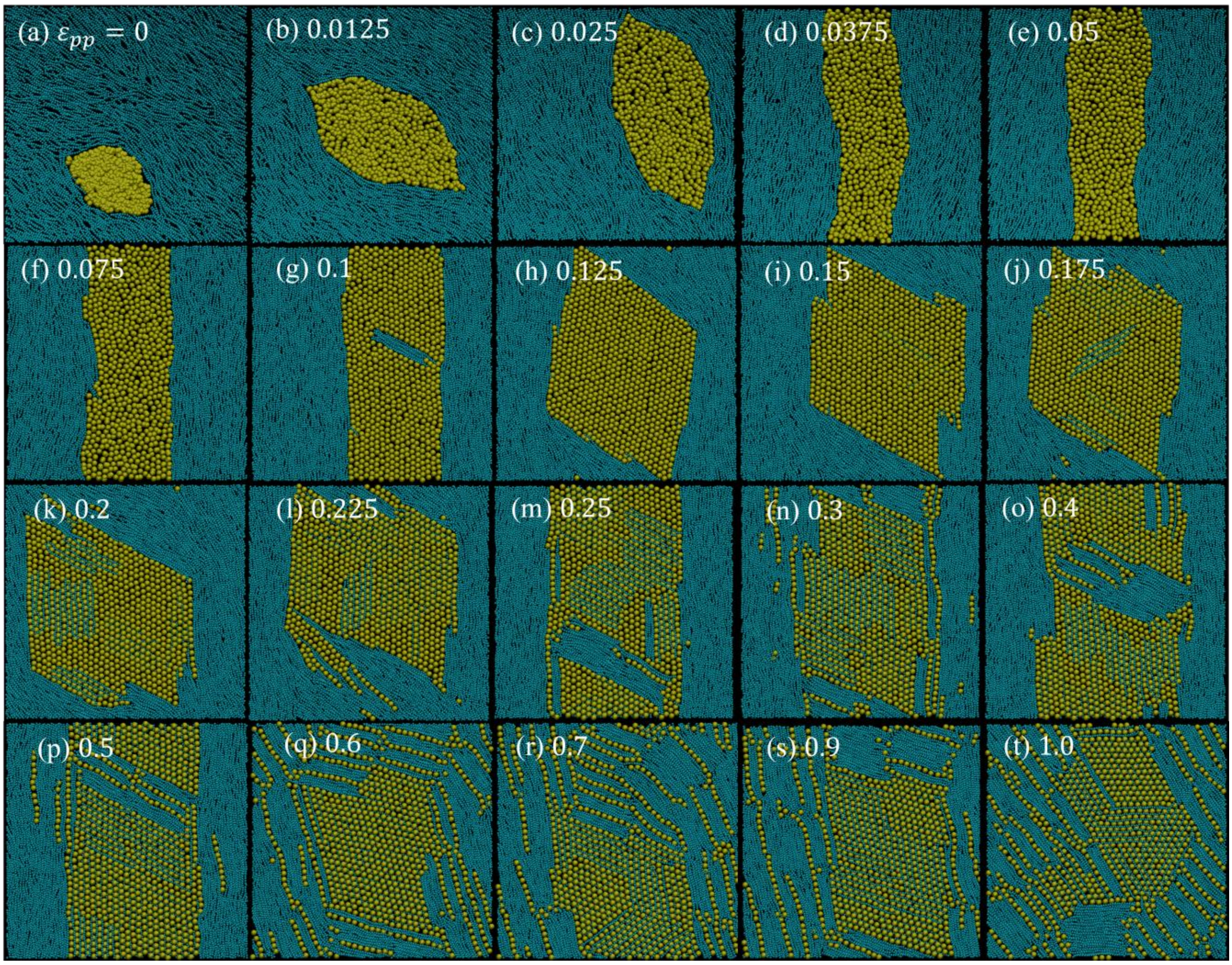

Figure 1: System configurations at the end of the Langevin simulations with different inter-NP repulsive strengths, $\varepsilon_{pp}$ (values shown in each panel). Cyan and yellow spheres represent monomers and NPs, respectively. The simulation cell contains 800 NPs and 500 rod-like chains, each having 25 monomers. The dimensions of the simulation cell are 100 × 100 × 4 along $x$, $y$ and $z$, respectively. Repulsive walls confine the $z$ dimension while periodic boundary conditions are applied on the $x$ and $y$ dimensions. Image created with the VMD software (version 1.9.3).[75]

As the interaction strength crosses a threshold, $\varepsilon_{pp} > 0.025$, the increasing repulsion destabilizes the droplet morphology, leading further to a reduction in NP overlaps (overlaps vanish for $\varepsilon_{pp} \geq 0.15$, see discussion in Section 3.3). The NPs, however, continue to form a single cluster, which becomes progressively more elongated.

Eventually, this combination of size increase and shape change makes the NP aggregate percolative, spanning the opposite sides of the simulation cell. This morphological alteration triggers a transition from the droplet to the slab form. The slab morphology is energetically favorable here as, combined with the non-bonded interactions, it does not necessitate the bending of polymers to enclose the NP-rich domain. As the repulsion increases further $(0.025 < \varepsilon_{pp} \leq 0.1)$, the width of the slab gradually increases (Figure 1(d)-(g)), accompanied by emergence of crystalline order among NPs for $\varepsilon_{pp} = 0.1$.

As the inter-NP repulsion becomes even stronger $(0.1 < \varepsilon_{pp} \leq 0.225)$, the slab configuration transforms into a rhombohedral NP droplet of high local order, as can be seen in Figure 1(h)-(l). The quantitative analysis based on the CCE metric, to be presented in section 3.3, supports the emergence of crystalline domains, as revealed here by visual inspection. For $\varepsilon_{pp} \geq 0.175$, polymer chains penetrate the still ordered NP, marking the onset of droplet fragmentation. Our observations suggest that extensive NP crystallization occurs at an intermediate intensity, as the droplet domain remains sufficiently large to accommodate the formed crystals while the np-np repulsions are still weaker than the NP-polymer ones. As repulsion increases further ($\varepsilon_{pp} > 0.225$) the droplet domain disperses progressively, with the largest of its fragments exhibiting high level of local ordering. Concomitantly, polymers form smectic domains which grow in extent with increasing NP repulsion. The smectic domains emerge because the volume occupied by the NP-rich domains keeps expanding, thus reducing the volume available for polymers. The enhanced confinement promotes polymer ordering into a smectic phase, resembling a lyotropic-like transition, driven by the increase in effective polymer concentration[33,54,55,76].

### *3.2.1. Self-Organization: Cluster Analysis*

The final snapshots, shown in Figure 1, point to self-organization of the system depending on the interaction strength between NPs. This self-organization spans different length scales: First, phase separation, at the global level, which splits the system into NP-rich and polymer-rich domains, which can form a well-defined interface or penetrate each other. Then, polymer chains can align and orient to create nematic and smectic phases. Finally, NPs can crystallize locally to create highly defected or well-shaped ordered morphologies. To further quantitatively characterize self-organization, the established morphologies and the corresponding transitions, we perform a cluster-based analysis for the full range of $\varepsilon_{pp}$ focusing on NPs. Towards this we adopt the geometric algorithm for cluster detection, as implemented in the Simu-D software[70], which has been employed in the past to gauge aggregation and crystal formation of attractive polymers[71], and caffeine clustering and stacking in membrane penetration[72]. The threshold radius for the cluster identification, $r_c$, is here set at $r_c = 1.1\sigma_p$ ($r_c = 2.75\sigma_m$). This cutoff distance criterion corresponds to the nearest neighbors of the first shell, as will be explained in the continuation.

The output of the cluster analysis includes the number of distinct clusters, $n_{clusters}$, ($n_{clusters} \epsilon \left[1, N_p\right]$), the size of cluster $i_c$, $N_{cl}(i_c)$, ($N_{cl}(i_c) \epsilon \left[1, N_p\right]$ and $i_c \epsilon [1, n_{clusters}]$) the number of distinct clusters with size larger than 1 (excluding practically isolated clusters formed by a single NP), $n_{clust}$, ($n_{clust} \epsilon \left[0, \frac{N_p}{2}\right]$), the cluster size distribution, and the size of the largest cluster, $N_{largest}$, ($N_{largest} \epsilon \left[1, N_p\right]$). In all the above, size corresponds to the number of NPs that form a cluster. The parameters $n_{clusters}$, $n_{clust}$, $N_{cl}(i_c)$ and $N_{largest}$ can be trivially normalized by the number of NPs, $N_p$, into normalized cluster count, $\rho_{clusters} = n_{clusters}/N_p$, normalized cluster count with size larger than 1, $\rho_{clust} = n_{clust}/N_p$, normalized size of cluster $i_c$, $v_{cl}(i_c) = n_{cl}(i_c)/N_p$, and normalized size of the largest cluster, $v_{largest} = N_{largest}/N_p$.

Additionally, the shape measures of the largest cluster are provided: asphericity, $b$, acylindricity $c$, and relative shape anisotropy, $k^2$. Towards this we calculate the analogue of the moment of inertia tensor considering that the largest cluster is composed of NPs of unit mass. The shape measures and the trace, $tr$, are extracted from the eigenvalues of the tensor built from the position vectors of the NPs with respect to the center of mass of the cluster. The asphericity and acylindricity reported here are rendered dimensionless by dividing by half the trace of the tensor: $2b/tr$ and $2c/tr$. The exact procedure for the shape measure analysis, including their definitions, can be found in Appendix A.3 of Herranz *et al.*[77] and in Bouzid *et al*[78]. The radius of gyration of the largest cluster is also calculated as[79]:

$$R_g = \sqrt{\frac{\lambda_1 + \lambda_2 + \lambda_3}{2N_{largest}}} \tag{9}$$

where $\lambda_1$, $\lambda_2$ and $\lambda_3$ are the eigenvalues of the tensor, ranked in descending order $\lambda_1 \geq \lambda_2 \geq \lambda_3$.

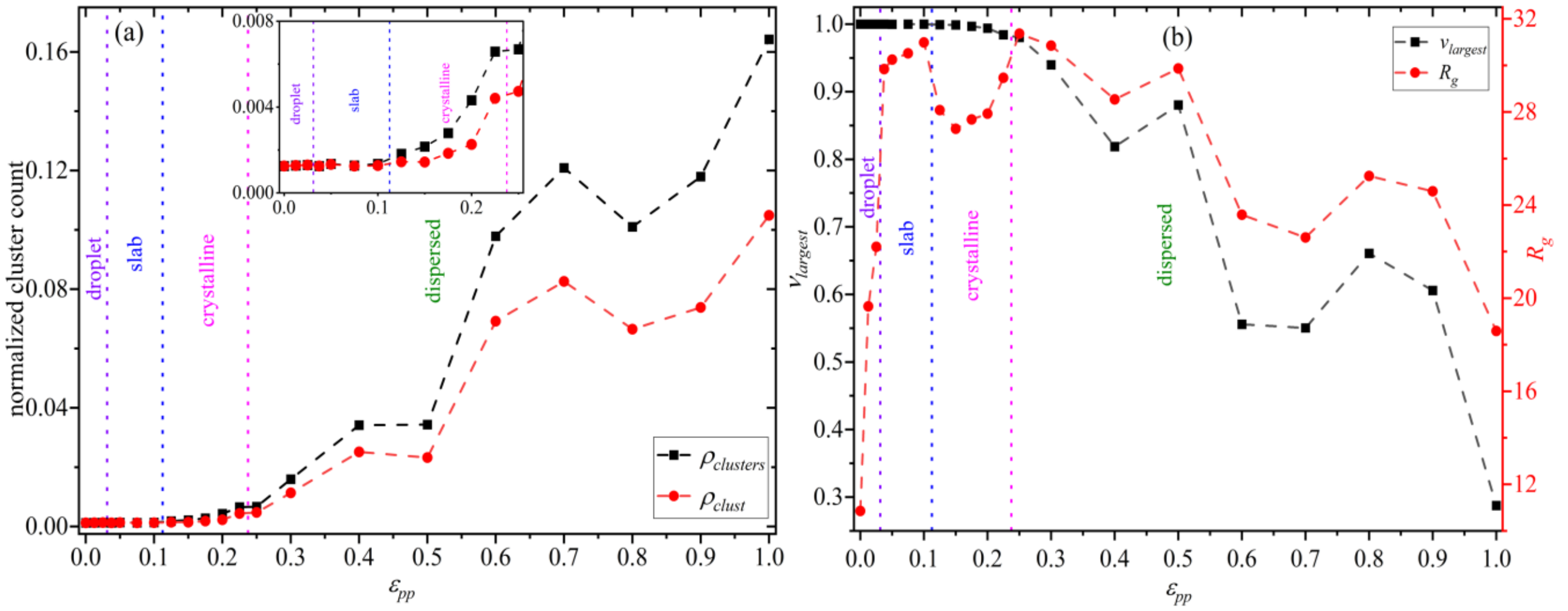

Figure 2. Results of the cluster analysis performed for the NPs on the equilibrated part of the simulation trajectory. (a) Normalized cluster count, $\rho_{clusters}$, (black squares) and its variant considering only clusters with size > 1, $\rho_{clust}$, (red circles) as a function of repulsion intensity between NPs, $\varepsilon_{pp}$. Inset: same as in the main panel but focusing on the low-intensity range. (b) Normalized size of the largest cluster, $v_{largest}$, (black squares, left $y$-axis) and radius of gyration of the largest cluster, $R_g$, (red circles, right $y$-axis) as a function of $\varepsilon_{pp}$. Vertical dotted lines identify the crossovers between different NP morphologies denoted as "droplet", "slab", "crystalline" and "dispersed". Dashed lines connecting the data points serve as a guide for the eye.

Panels (a) and (b) of Figure 2 and Figure 3 summarize the main findings of the cluster analysis. Vertical lines indicate the onset of the three transitions between the different morphological regions with increasing strength as identified here: droplet → slab (i → ii), slab → crystalline droplet (ii → iii), crystalline droplet → dispersed (iii → iv). Panel (a) presents the dependence of the number of normalized cluster count, $\rho_{clusters}$, and of its variant $\rho_{clust}$, on the repulsion strength, $\varepsilon_{pp}$. The data quantitatively supports the observed self-organization, the established morphologies and the corresponding conformational changes. In the low repulsion regime ($0 \leq \varepsilon_{pp} \leq 0.15$), the NPs remain inter-connected as a single cluster, as indicated by the combination of $v_{largest} = 1$ and $\rho_{largest} \rightarrow 1/N_p$, consistent with the snapshots in Figure 1 (a-i). Within this range, which spans effectively the droplet, slab and the early part of the crystalline droplet regimes, the cluster, composed of all NPs, retains its structural integrity. While a very small fraction of polymers occasionally infiltrates the NP domain for $0.1 \leq \varepsilon_{pp} < 0.2$, the NP domain remains sufficiently connected to form a single cluster.

For $\varepsilon_{pp} \geq 0.2$, deep in the crystalline droplet regime, the NP-rich domain progressively fragments and disperses into multiple clusters, as reflected by the monotonic increase in the number of distinct clusters (panel a). This is also signaled in the normalized size of the largest cluster which progressively drops from the initial value of unity. This dispersion can be largely divided into three stages. For $0.2 \leq \varepsilon_{pp} \leq 0.25$, the dispersion proceeds gradually, characterized by a slow but steady increase in both $\rho_{clust}$ and $\rho_{clusters}$, and a simultaneous decrease in $v_{largest}$. This marks the onset of fragmentation: while the main cluster starts losing NPs, its integrity remains very high, as indicated by $v_{largest} \approx 0.98$ at $\varepsilon_{pp} = 0.25$. Then, an intermediate regime $0.25 < \varepsilon_{pp} \leq 0.5$ is characterized by a systematic and more intense fragmentation while the single cluster remains the dominant NP morphology. At even larger repulsion strengths $0.5 < \varepsilon_{pp} \leq 1.0$, fragmentation is completed, as indicated by the sharp decline in the size of the largest cluster and the analogous increase in the normalized cluster count. The cluster statistics become highly prone to statistical noise at higher intensity values due to the finite size of the system along with the applied confinement on the $z$ dimension and the thermal fluctuations: smaller clusters merge to form larger ones, which in turn dissolve to form smaller ones. However, the general trend is effective mixing and NP dispersion which are evidenced by $\rho_{clust}$ marking its maximum value ($\rho_{clust} \approx 0.164$) and $v_{largest}$ registering its minimum value ($v_{largest} \approx 0.29$) at the highest NP repulsion studied here, $\varepsilon_{pp} = 1$. Given the system size and the wall confinement, full dispersion is not achievable, as reflected by the numbers cited above which deviate significantly from the limit (ideal) values of $\rho_{clusters}^{max} = 1$ and $v_{largest}^{min} = 1/N_p$ (= 1.25×10$^{-3}$). The cluster trends can be explained as higher repulsion strength forces the NP assembly to expand, leading to greater polymer infiltration (and eventual mixing) and to the formation of small, isolated clusters that progressively get separated from the main one.

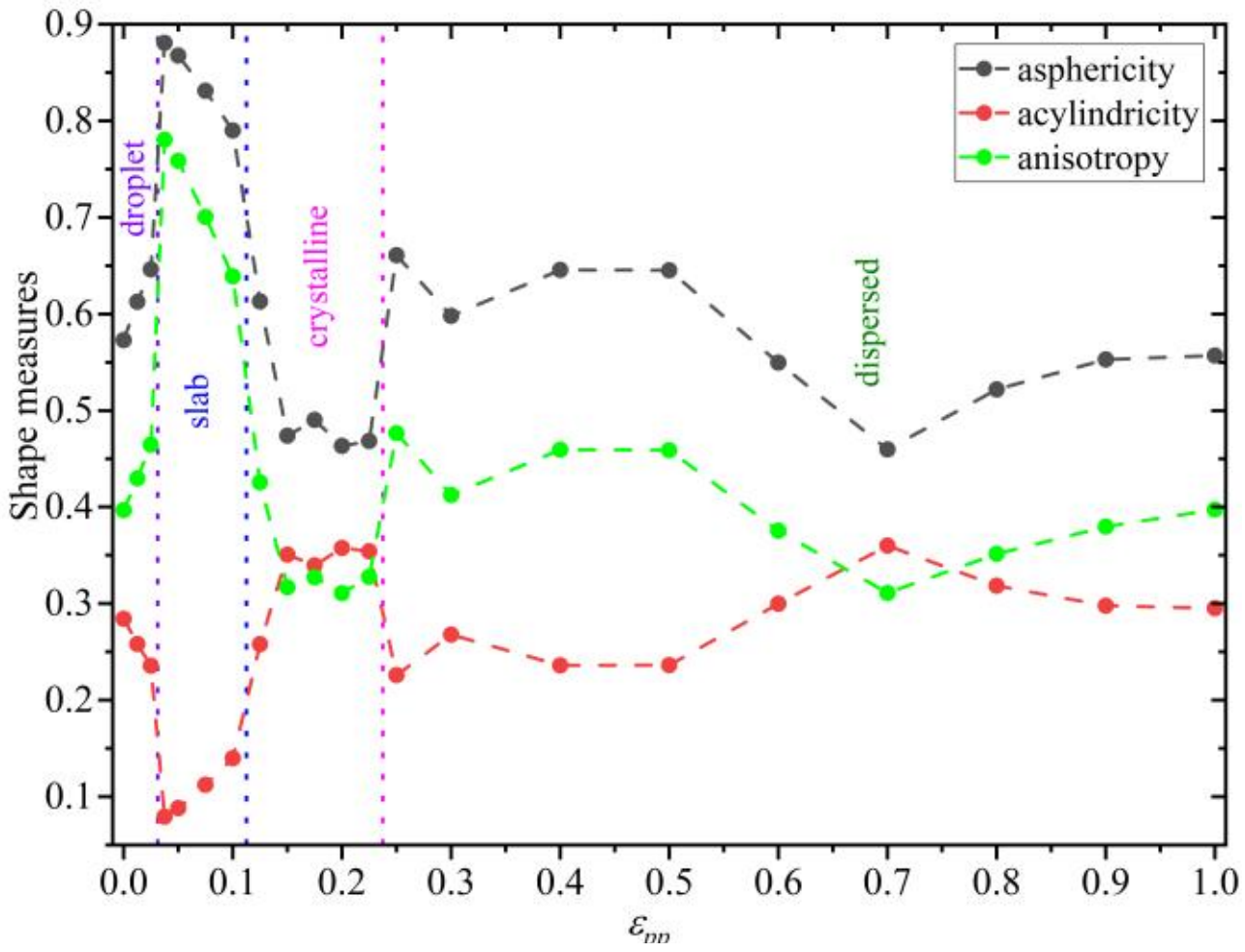

Figure 3. Shape measures as obtained from the mass-moment of inertia tensor for the largest cluster of NPs: (reduced) asphericity, (reduced) acylindricity and relative shape anisotropy as a function of $\varepsilon_{pp}$. Vertical dotted lines identify the crossovers between different NP morphologies denoted as "droplet", "slab", "crystalline" and "dispersed". Dashed lines connecting the data points serve as guide for the eye.

The radius of gyration of the largest cluster (red scattered points in panel (b) of Figure 2) echoes further the change in size of the NP assembly: initially the largest cluster adopts its most compact size, as mirrored by the minimum $R_g$ value. Once NP repulsion is activated the cluster expands rapidly in the droplet regime. Although the transition to the slab morphology slows down the cluster expansion, it remains systematic and reaches maximum value at the transition between slab and crystalline droplet. In this regime the non-trivial change in the cluster size is strongly related to the alterations occurring to the local environment, becoming more symmetric, to accommodate crystallization. The final transition (crystalline droplet to dispersed) is marked by an increase in the radius of gyration as the cluster, still intact before the gradual fragmentation, expands to reach its maximum size. The eventual dispersion and collapse of the principal NP cluster lead to systematically smaller size; apart from the droplet phase, the minimum value for the radius of gyration corresponds to the maximum repulsion between the NPs ($\varepsilon_{pp} = 1$).

Figure 3 shows the evolution of the three geometric shape descriptors of the largest cluster as function of $\varepsilon_{pp}$. These shape measures are: (reduced) asphericity, (reduced) acylindricity, and relative shape anisotropy. For any 3D system, asphericity measures the deviation from spherical symmetry, acylindricity quantifies the deviation from cylindrical symmetry, while the shape anisotropy $k^2$, is a measure of overall anisotropy of the particle shape compared to isotropic shape (spherical), and is bounded in the interval [0, 1]. $b = 0$ for a sphere and $c = 0$ for a cylinder; the higher the deviation, the larger the value of $b$ and $c$. Rendering them dimensionless through a division by half the trace, $tr/2$, further bounds them in the interval [0, 1]. In the analysis of the shape measures we should further consider that extreme confinement is applied along $z$, restricting the shape of the cluster from adopting specific configurations as it would in the 3D bulk case.

In the droplet phase i.e. in the low $\varepsilon_{pp}$ regime ($0 \leq \varepsilon_{pp} \leq 0.025$), asphericity and relative shape anisotropy both increase gradually with $\varepsilon_{pp}$, while acylindricity drops at a very similar rate. Based on this, the NP cluster becomes less spherical, less symmetric and more cylindrical. It is worth noting that the quasi-2D confinement strongly flattens the droplet, restricting its dimension along the slit width to a much smaller value than its two lateral dimensions within plane. Consequently, the smallest principal eigenvalue, associated with the droplet width, remains much smaller and nearly constant compared to the two larger eigenvalues corresponding to its lateral extent. This intrinsic eigenvalue disparity keeps the anisotropy and acylindricity relatively high throughout this regime. With increasing $\varepsilon_{pp}$ the droplet adopts a prolate shape. becoming progressively more elongated along the confined $z$ direction. This is reflected by a decrease in acylindricity and an analogous increase in asphericity and shape anisotropy.

The droplet → slab transition is marked by a sharp change in all three shape measures of the largest cluster. At $\varepsilon_{pp} = 0.0375$, which marks the onset of the slab phase, asphericity and anisotropy jump sharply from 0.65, and 0.46 to their maximum values ($\sim 0.85$ and $\sim 0.75$, respectively), while acylindricity conversely decreases from ~0.24 to its global minimum ($\sim 0.09$). These abrupt changes are attributed to the shape transformation of the cluster from droplet to slab. The cluster spans two opposite sides of the simulation cell, forming a rectangular slab, whose length and breadth contribute to the largest and intermediate eigenvalues, respectively. The pronounced asymmetry between these two dimensions leads to the sharp rise in asphericity and anisotropy. Visual inspection of the snapshots in Figure 1 verifies the transition from a spherical-like cluster to an elongated, cylindrical-like assembly of NPs. Upon further increasing $\varepsilon_{pp}$, still within the slab phase ($0.0375 \leq \varepsilon_{pp} \leq 0.1$), the breadth of the slab increases, while the length remains constant resulting in a gradual decrease in asphericity and anisotropy and an increase in acylindricity.

The slab → crystalline NP droplet transition is again accompanied by a very sharp decrease in asphericity and relative shape anisotropy and a steep increase for acylindricity. These changes arise from the loss of the percolating slab conformation and the successive stabilization of the NP cluster into a rhombus-shaped crystalline droplet, as can be seen in the snapshots given in Figure 1 for $\varepsilon_{pp} = 0.15$. The rhombus-shaped droplet, whose largest and intermediate eigenvalues are similar, causes the observed decrease in both asphericity and anisotropy. NP crystallization involves a shift in shape; the largest NP assembly becomes more spherical and symmetric due to analogous structural changes toward an ordered state at the local level. In the crystalline droplet regime, the shape measures remain constant, with minimal fluctuations, as a function of repulsion. The final transition to dispersed state is marked by an increase in anisotropy and asphericity. While the principal cluster decomposes progressively with increasing intensity its shape is not significantly altered. Large fluctuations, due to thermal fluctuations, are observed as cluster segments or individual NPs get detached or merged into the largest cluster. It should be noted that the physical concept of the largest cluster is qualitatively different between, for example, the dispersed and the droplet phases. In the former, it is composed of a smaller group of NPs so limited additions or removals could lead to significant fluctuations in shape and size; in contrast, the droplet-like cluster contains all NPs, adopts a compact size and remains stable, within small thermal variations.

### *3.2.2. Self-Organization: Voronoi Analysis*

Tuning the inter-NP repulsion by varying $\varepsilon_{pp}$ affects the overlapping between the NPs, leading to a change in the local density. To examine the local packing, we perform Voronoi tessellation. Towards this we use the voro++ software[80], implementing periodic boundary conditions in dimensions $x$, $y$ and confinement in $z$. Figure 4(a) presents the average Voronoi volume around each NP, $V_p$ and monomer, $V_m$, as a function of repulsion strength. Given that the Voronoi tessellation is space-filling, it stands $MN_mV_m + N_pV_p = L_xL_yL_z$, identifying an inverse correlation between $V_m$ and $V_p$ which is clearly reflected in the trends observed in Figure 4(a).

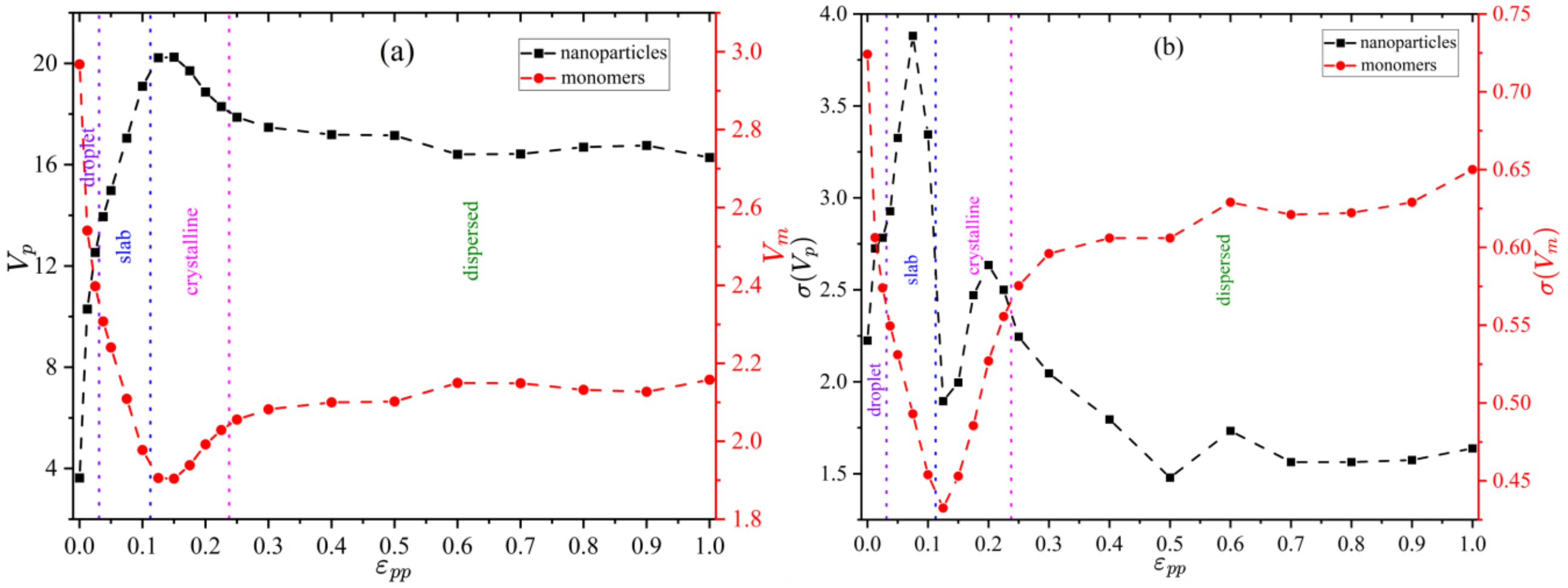

Figure 4. (a) Average Voronoi volume for NPs (black squares, *y*-axis), $V_p$, and for monomers (red circles, right *y*-axis), $V_m$, as a function of $\varepsilon_{pp}$. (b) Variance of the Voronoi volume for NPs (black squares, *y*-axis), $V_p$, and for monomers (red circles, right *y*-axis), Vm, as a function of $\varepsilon_{pp}$. Vertical dotted lines identify the crossovers between different NP morphologies denoted as "droplet", "slab", "crystalline droplet" and "dispersed". Dashed lines connecting the data points serve as guide for the eye.

Monomers and NPs exhibit their most dilute and dense environments, respectively, for $\varepsilon_{pp}$ = 0. This is not surprising given that the ghost interactions between NPs and the high repulsion between NPs and monomers (and between monomers themselves) lead to the formation of a compact cluster of extended overlaps. As repulsion intensity increases, the cluster expands and the local environment for NPs becomes progressively more diluted as the total volume of the simulation cell remains constant. Upon entering the slab morphology, the change in the local environment remains consistent and systematic but the rate of changes drops compared to the initial droplet regime. The transition from slab to crystalline droplet ($\varepsilon_{pp} \geq 0.125$) is marked by the Voronoi volume registering its global maximum for the NPs and its corresponding global minimum for the monomers. This trend is further accompanied by a peak in the local order as NPs tend to form quasi-2D crystals in the interior of the formed droplet. While this may appear as a paradox the establishment of dilute environment, combined with a more symmetrical shape and size of the main cluster allows for the NPs to adopt positions which are compatible with the triangular (TRI) crystallites, as will be demonstrated in Section 3.3. For example, at $\varepsilon_{pp}$ = 0.125 and 0.15, the average Voronoi volume for NPs reaches its maximum value (~20.2), while the crystallinity of the system peaks at $\varepsilon_{pp}$ = 0.15. In parallel, the local environment around the polymer segments becomes necessarily denser. As interaction intensity between NPs increases further, the NPs disperse and $V_p$ decays until reaching a plateau value at $\varepsilon_{pp}$ ~ 0.6. Consequently, $V_m$ also stabilizes after a progressive increase in the crystalline droplet and early dispersed regimes. These stable values indicate that both polymers and nanoparticles have reached a stable local packing density.

Figure 4(b) shows the variance as calculated for the distribution of the Voronoi volume for NPs and monomers. With respect to the latter, maximum deviation is observed at zero interaction strength, followed by a systematic decrease for the whole range of repulsion in the droplet and slab regimes. The minimum in deviation for the Voronoi volume of monomers corresponds to $\varepsilon_{pp}$ = 0.125. In contrast, the variance of the Voronoi volume for the NPs increases continuously until reaching a global peak in the slab regime, which further corresponds to the regime of the lowest local order. A sharp drop is then observed during the slab → crystalline transition. In the first three regimes, the distributions of the Voronoi volumes become the narrowest for both the nanoparticles and the polymer segments as the system transits to the ordered droplet morphology. Through the perspective of crystallization, to be presented in the continuation, the maximum of local order for NPs corresponds to the minimum in the variance of the local volume. Effectively, the NPs are positioned in regular patterns dictated by the lattice sites of the crystal, and their local environment is highly uniform. Finally, in the dispersed phase, where NPs are in contact with the polymer segments, the corresponding mixing leads to a more uniform local environment for the NPs while the inverse happens for the monomers. Beyond $\varepsilon_{pp} \approx 0.6$, the variance of the Voronoi volume for both species reaches a plateau, which indicates that no significant structural changes take place in the system.

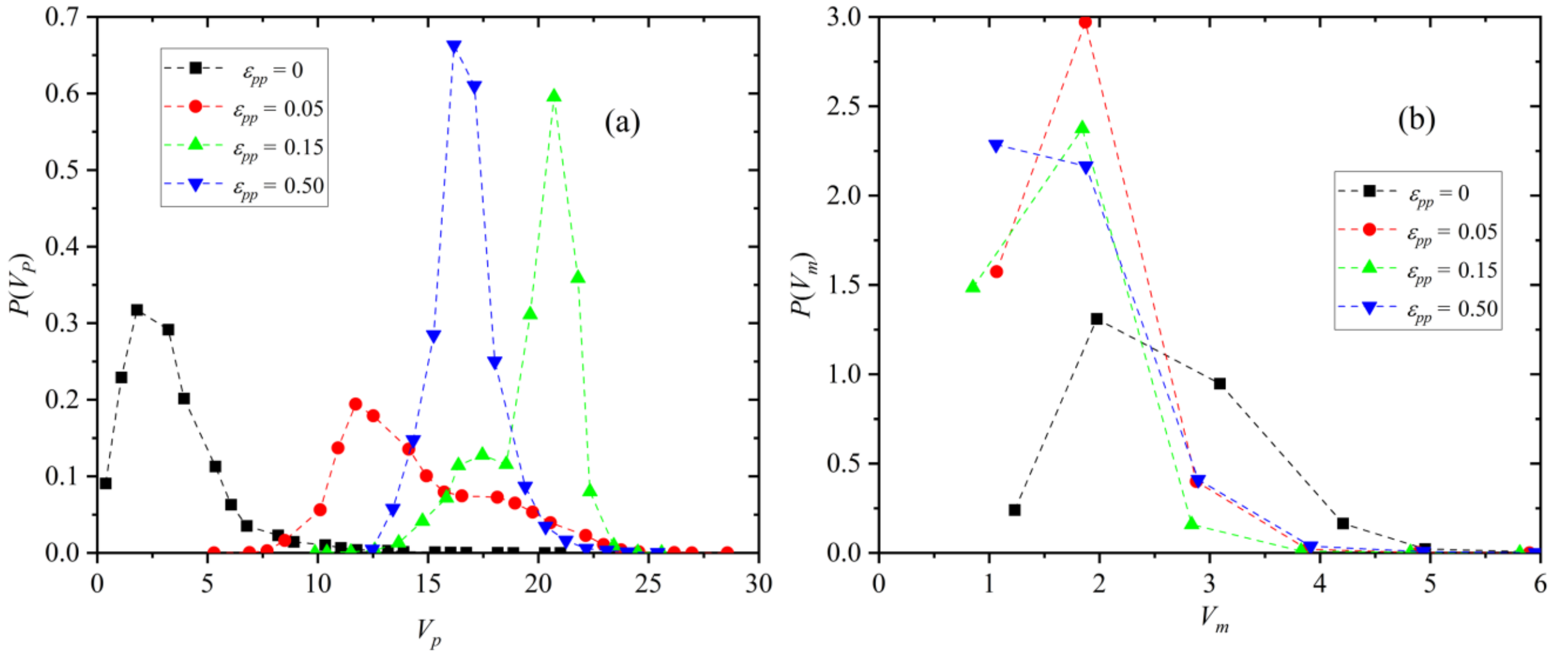

Figure 5. Probability distribution function for the Voronoi volume around NPs (panel a) and monomers (panel b) for different $\varepsilon_{pp}$ values.

Panels (a) and (b) in Figure 5 host the probability distribution function of the Voronoi volume for NPs and monomers, respectively. For NPs at zero repulsion (droplet phase) the Voronoi volume follows a Boltzmann-like distribution with a peak around very low values ($V_P \sim 2.5$), indicative of very dense local packing within the compact NP cluster, where extensive overlapping between NPs occurs. The tail towards the higher volume is attributed to the NPs belonging to the external surface of the cluster, which are in contact with the monomers of the polymer chains. Activating even low repulsion ($\varepsilon_{pp} = 0.05$, slab phase) shifts the distribution to higher values, corresponding to more dilute local environment. The distribution becomes broader, showing a primary peak at $V_P \sim 12$ and an inter-connected one at ~ 18. The more pronounced peak at the lower volume is associated with NPs in the interior of the cluster, whereas the weaker second peak arises from the NPs residing on the NP-monomer interface. The corresponding variance is considerably high, as also confirmed by the results in Figure 4(a).

As the repulsion increases further to $\varepsilon_{pp} = 0.15$ (crystalline droplet regime) a sharp peak appears at $V_p \sim 21$, overlapping with a less intense, and relatively diffused peak at, $V_p \sim 18$. The dominant peak at higher $V_p$ can be attributed to the crystalline NPs residing in the interior part of the droplet showing a highly homogeneous and ordered local environment. In contrast, the weaker peak corresponds to the NPs lying on the interface, which additionally have a denser environment around them. This is because polymer segments, by being smaller can pack efficiently next to them. At even higher repulsion ($\varepsilon_{pp} = 0.5$), the Voronoi distribution of the NPs becomes very narrow and symmetric following a single Gaussian-like behavior. Such homogenous packing density is caused by the NP-monomer mixing which is favored at this repulsion range. Effectively, this mixing ends the separation of NPs into "interior" ones, that have other NPs as nearest neighbors, and "interfacial" ones which lie adjacent to both other NPs and polymer segments.

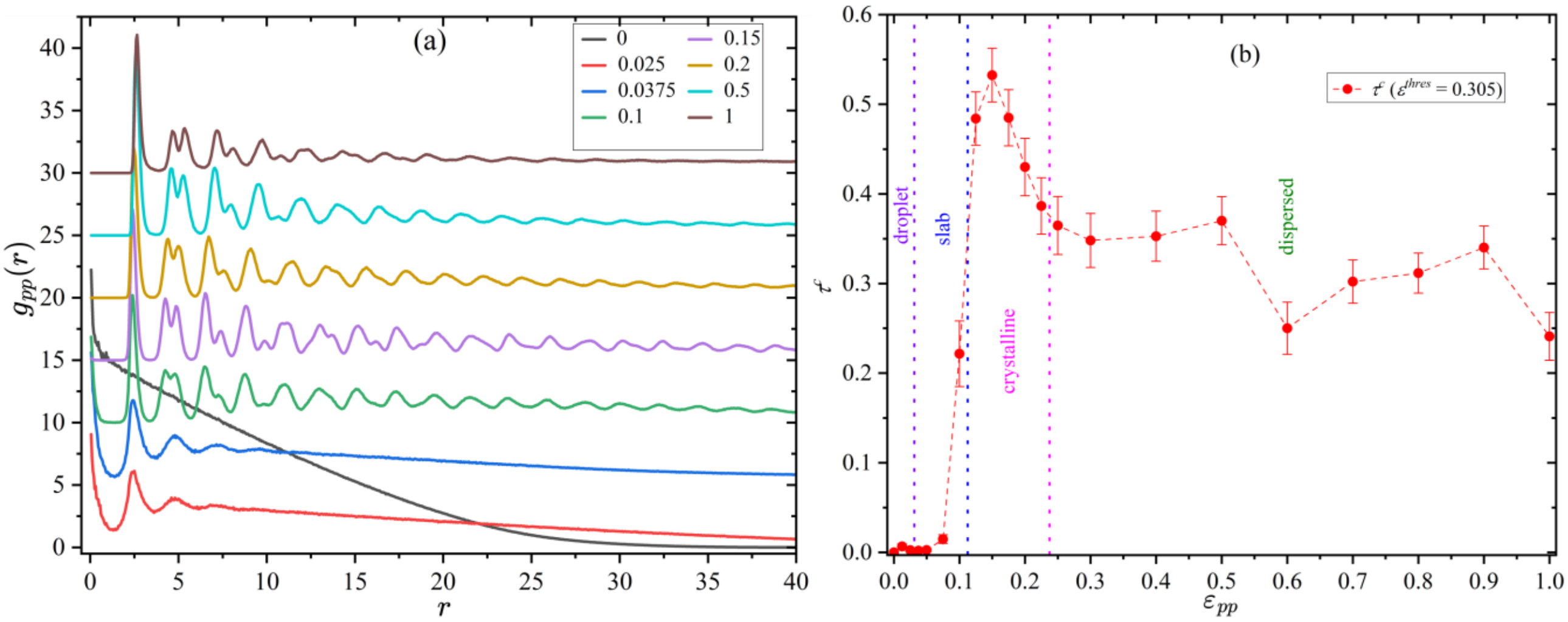

Figure 6. (a) Radial distribution function of NP pairs, $g_{pp}(r)$, as a function of distance, $r$, for different repulsion strengths. The $g_{pp}(r)$ curves for $\varepsilon_{pp} \geq 0.0375$ have been shifted vertically by increments of 5 units for visual clarity. (b) Crystallinity, as a function of repulsion strength, $\varepsilon_{pp}$. Vertical dotted lines identify the crossovers between different NP morphologies denoted as "droplet", "slab", "crystalline" and "dispersed". Dashed lines connecting the data points serve as guide for the eye.

With respect to monomers (Fig. 5(b)), except for the zero-repulsion limit, where the distribution is wider and the environment is less dense, all distributions in the other three regimes appear quite similar in shape. This weak dependence on $\varepsilon_{pp}$ is indicative of a less pronounced effect of the NP repulsion on the local packing of the monomers. Furthermore, the bond and bending constraints due to chain connectivity, impose an inherent network of bonded neighbors for each monomer: an end segment is connected to a single monomer, while an inner chain segment is connected to two bonded beads. Thus, it is expected that these bonded interactions, along with the significantly larger population of monomers compared to NPs, renders the corresponding distribution of the local packing density narrower and less prone to significant alterations than that of NPs.

### *3.3. Self-Organization: NP Crystallization*

Due to the imposed strong confinement ($L_z = 4$), the NPs with a diameter of 2.5 essentially form a thin quasi-2D layer at the center of the slit, as evident from Figure S1 (NP density profile along z). Based on this arrangement, the 2D-pair radial distribution function for the NP pairs, $g_{pp}(\rho)$ is calculated. For this calculation, only the pairs satisfying the condition $|\Delta_z \leq 1|$ are considered. $\Delta_z$ is the separation between the two NPs along the slit-normal ($z$-direction). Here, $\rho$ denotes the projected distance between two NPs in the $xy$ plane.

Panel (a) of Figure 6 depicts the pair radial distribution function between NPs, $g_{pp}(r)$, for different $\varepsilon_{pp}$. Zero repulsion intensity results in extensive overlaps between the NPs, leading to the formation of the compact droplet morphology, as evidenced by the snapshot of panel (a) in Figure 1, and as reflected by the global maximum in $g_{pp}(r)$ at $r \rightarrow 0$ for $\varepsilon_{pp}= 0$. $g_{pp}(r)$ decreases rapidly with distance and it drops to zero at approximately $r \approx 30$. This value marks the largest distance separating NPs in the formed aggregation. As repulsion intensity increases the fraction of overlapping NPs decreases significantly as evidenced by the sharp drop in the $g_{pp}$ in the overlapping regime $\left(0 \leq r \leq \sigma_p\right)$. Apart from the extensive overlaps the pair radial distribution for $\varepsilon_{pp} \leq 0.0375$, belonging to the droplet and slab regimes, reveals a disordered NP packing without any traces of short- or long-range order. For $\varepsilon_{pp} = 0.1$, near the end of the slab regime, the number of overlaps is reduced, and a global maximum occurs at NP contact: $r = \sigma_p = 2.5$. Additionally, regular patterns in the $g_{pp}$ reveal the onset of early crystallization which coincides with the transition from the slab to the crystalline droplet. Importantly, all systems except $\varepsilon_{pp} = 0$ show convergence to $g_{pp}(r) \approx 1$ at long distances $(r > 35)$, consistent with sufficiently dispersed systems where the probability of finding a NP at large separations approaches the bulk average density.

Once the employed repulsion strength reaches a value of $\varepsilon_{pp} = 0.15$, corresponding to the crystalline droplet regime, the short-range overlaps are eliminated and the first peak of $g_{pp}(r)$ becomes progressively more intense. Regularity in the appearance of maxima and minima, beyond the nearest and second-nearest neighbors clearly identifies the presence of a crystal morphology. Notably, at $\varepsilon_{pp}$= 0.15 the peaks are the sharpest, and persist over the longest range, indicating a long-range crystalline structure for the NPs. The positions of the first three peaks are approximately at 2.5, 4.33 and 5, which when normalized by the NP diameter, correspond to 1, 1.732 $(\sqrt{3})$ and 2, geometric distances compatible with a 2D triangular (hexagonal closed-packed) arrangement.

The trends observed in panel (a) of Figure 6 clearly demonstrate that a transition to crystal morphology takes place for NPs once a critical repulsion strength is reached. This is further confirmed by the regularity of the NP packings as observed for example in panels (g)-(j) of Figure 1. To gauge and quantify such a phase transition we employ the Characteristic Crystallographic Element (CCE) norm[73,74]. The CCE norm algorithm is a refined metric of local order which utilizes crystallographic operations, actions and elements to quantify the similarity of a computer-generated local environment with respect to reference crystals in two and three dimensions, in the bulk and under confinement[81,82]. Details of the methodology and its practical implementation for a series of different physical system can be found in Ref.[73,74]. In the present study, extreme confinement is applied on the z dimension, given that the NP size ($\sigma_p$ = 2.5) is quite close to the inter-wall distance ($L_z$ = 4), leading to quasi monolayer films with respect to NPs. Thus, we consider the *x-y* projections of the coordinates of the NPs, and we gauge the similarity of the local environment around each NP with respect to the honeycomb (HON), square (SQU) and triangular (TRI) crystals and the pentagonal (PEN) local symmetry in two dimensions as in the past studies on 2D packings of hard-sphere polymers[81,82]. A CCE norm, $\varepsilon^X(i)$, is obtained for each NP, *i*, against every reference crystal (and local symmetry), *X*, in two dimensions. The lower the value of the CCE norm, $\varepsilon^X(i)$, the higher is the similarity of the given local environment, *i*, to that reference crystal *X*. Once the CCE norm is below a threshold, $\varepsilon^{thres}$, ($\varepsilon^X(i) \leq \varepsilon^{thres}$), which is commonly set between all reference crystals the local environment of particle, *i*, is considered of having *X* similarity. By construction for a given site *i*, the CCE norm value cannot be simultaneously too low with respect to two different crystals *X* and *Y* ($\varepsilon^X(i) \rightarrow 0$, $\varepsilon^Y(i) >> 0$ and vice versa: $\varepsilon^Y(i) \rightarrow 0$, $\varepsilon^X(i) >> 0$). The analysis is repeated over all particles and reference crystals for a given system configuration (frame) and over all equilibrated frames in the simulation trajectory. This allows for the calculation of the CCE order parameter for each reference crystal, $S^X$ ($S^X \in [0, 1]$), and for the total degree of crystallinity, $\tau^c$ ($\tau^c \in [0, 1]$), calculated as the sum of all order parameters ($\tau^c = S^{\mathrm{SQU}} + S^{\mathrm{HON}} + S^{\mathrm{TRI}}$). The pentagonal local symmetry is expected to act as a competitor to crystallization in two dimensions[81], like the fivefold in three dimensions. Accordingly, the calculation of the PEN CCE order parameter constitutes part of the present crystallographic analysis.

For non-overlapping hard sphere systems, the CCE norm is normalized by the characteristic length of the system, $l_{CCE}$, which coincides with the sphere diameter (see for example Eqs. 2, 5 and 6 in Ref.[74]). A similar approach is adopted in the case of fussed Lennard-Jones (LJ) spheres, considering the LJ collision diameter[83]. In the present study the situation is different as the variation of the repulsion strength results in diverse configurations, ranging from the ideal-gas limit for very low values of repulsion strength to the hard-sphere limit for higher intensities. This is evidenced by the behavior of $g_{pp}(r)$ as seen in panel (a) of Figure 6. To account for this in calculation of the CCE norm and the corresponding crystallinity, we calculate three different variations corresponding to $\varepsilon^{thres}$ = 0.245, $l_{CCE} = \sigma_p$ = 2.5; $\varepsilon^{thres}$ = 0.305 and $l_{CCE}$ = 2.5; $\varepsilon^{thres}$ = 0.305, $l_{CCE}$ = 2.5, $r_{min} = 3.75$ and $\tau^{c,mod} = n^{CCE}\tau^c$. The third variant includes a normalization factor which multiples the degree of crystallinity of the form:

$$n^{CCE} = \int_0^{1.5\sigma_p} \frac{g_{pp}(r,0)dr}{\int_0^{1.5\sigma_p} g_{pp}(r,\varepsilon_{pp})dr} \tag{10}$$

It is employed to account for the reduced distances, and thus the lower CCE norm values for the systems whose low repulsion intensity allow for overlaps between the NPs, primarily in the initial droplet and secondarily in the slab regimes.

Another important point for consideration in the calculation of the order parameters and of the crystallinity is that the NPs that belong to the NP-monomer interface do not have sufficient NP nearest neighbors for the crystallographic analysis. Their Voronoi polygon is incomplete, resulting in a very high CCE norm and accordingly a very low similarity to any of the reference crystals. Consequently, in the case of the droplet, slab or crystalline droplet regimes the NPs which lie on the NP-monomer boundary do not contribute to crystallinity and local ordering. In the final dispersed morphology due to extensive NP-monomer mixing crystallinity is thus expected to drop further as most of the NPs are surrounded by chain segments. In the present analysis no correction is implemented to account for the exclusion of the interfacial NPs from the crystallographic analysis.

Panel (b) of Figure 6 presents the results of the CCE-based analysis for $\varepsilon^{thres}$ = 0.305 and $l_{CCE}$ = 2.5, while Figure S3 in Supplementary material compares the trends between the three different variations utilized here. Fig S3 shows that all three variants yield the same regimes and the maximum crystallinity at $\varepsilon_{pp} = 0.15$. Accordingly, the discussion below focuses on the results presented in Figure 6(b) of the main text.

Figure S4 in Supplementary material shows the evolution of the CCE order parameters, $S^X$, and of the total crystallinity, $\tau^c$, as a function of timesteps from the LD simulation on the system with $\varepsilon_{pp}$= 0.15. For all systems that show crystallization the behavior is identical: the ordered morphology corresponds exclusively to triangular (TRI) crystals with no other ordered structures appearing in the final NP configurations. The level of the established crystallinity depends strongly on the applied repulsion intensity, as will be explained in detail in the continuation. Also, it is interesting to notice that the phase transition starts immediately once repulsion is activated and reaches the stable ordered morphology within a single frame ($2\times10^6$ timesteps).

In the droplet and slab regimes ($\varepsilon_{pp} \in$ [0.0125, 0.050]) the NP assembly is predominantly amorphous without traces of local order ($\tau^c \to 0$). It is only at $\varepsilon_{pp}$= 0.075 that the first isolated crystal nuclei are formed inside the NP slab, as indicated by the small increase in the degree of order ($\tau^c \approx 0.015$). At $\varepsilon_{pp}$= 0.10, a threshold value which marks the slab → crystalline droplet transition, a sharp increase is observed in the local order of the NP assembly ($\tau^c \approx 0.22$). The crystalline droplet regime is characterized by the highest crystallinity over the whole range of repulsion strength. The maximum achieved degree of ordering ($\tau^c \approx 0.53$) corresponds to $\varepsilon_{pp}$= 0.15, near the middle of the regime. Practically, more than half of the NPs possess a highly order local environment. Once the maximum is reached, and the cluster expands further due to the NP repulsion, the level of local ordering systematically drops. The transition to the dispersed regime is accompanied by a stabilization in crystallinity within large statistical fluctuations as clusters expand and shrink due to the thermal motion and mixing of NPs with polymer segments.

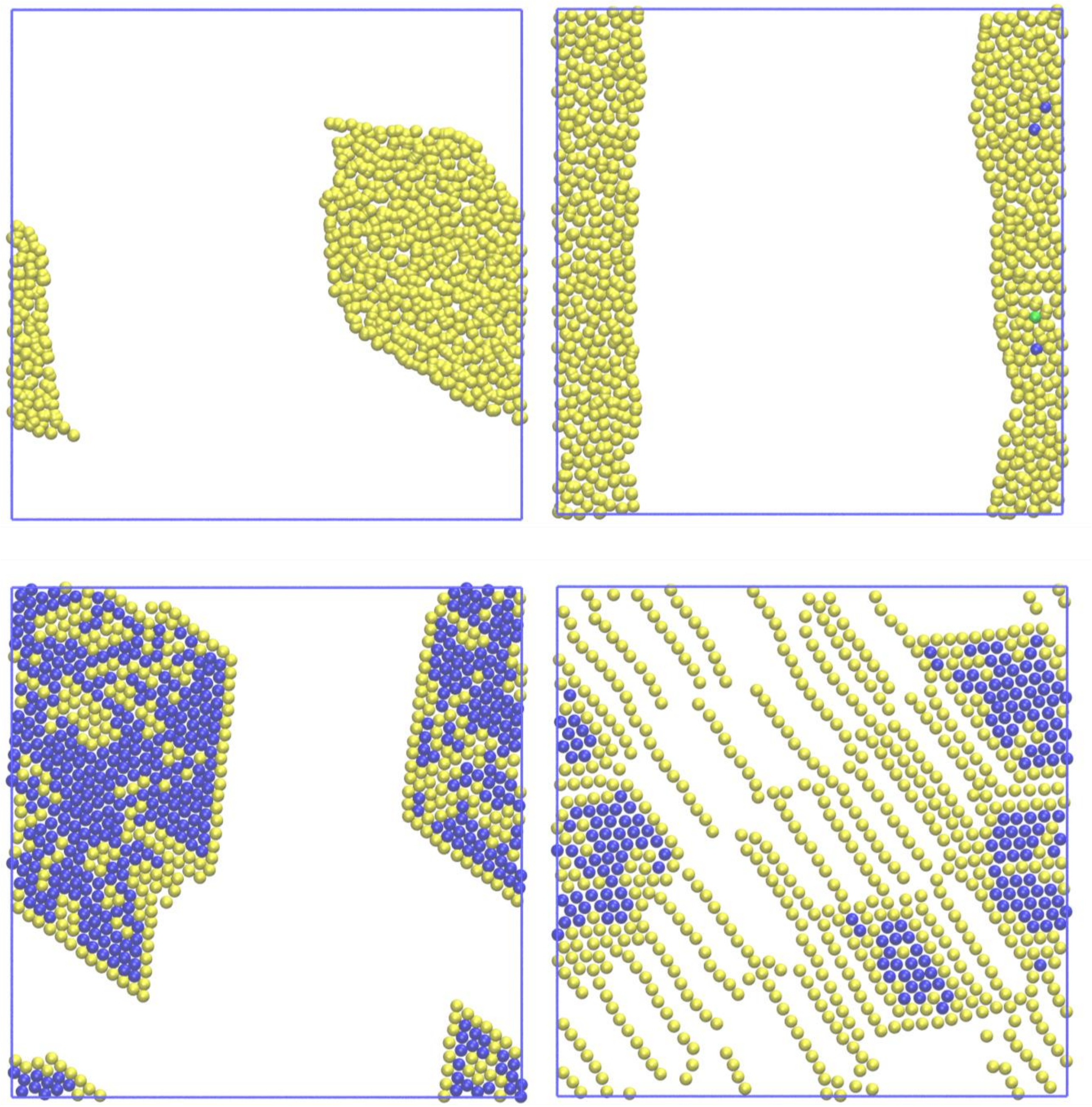

Figure 7. System configurations at the end of the Langevin simulations for $\varepsilon_{pp}$ = 0.0125 (top left), 0.05 (top right), 0.15 (bottom left) and 1 (bottom right). NPs are shown colored according to their crystal similarity as quantified by the CCE norm: blue and green colors correspond to NPs with triangular (TRI) and pentagonal (PEN) similarity, respectively. No sites with honeycomb (HON) or square (SQU) crystal symmetry are detected. Yellow color corresponds to NPs with no similarity to any of the reference crystals or the pentagonal local symmetry. For clarity, polymers are not included in the visualization. Image created with the VMD software[75].

The panels of Figure 7 show system configurations at the end of the LD simulations where NPs are colored according to their crystal or local symmetry similarity, as gauged by the CCE norm descriptor. Blue and green correspond to triangular (TRI) and pentagonal (PEN) similarity, respectively. For the presented configuration no NPs with honeycomb (HON) or square (SQU) symmetry are detected. Yellow color corresponds to amorphous (AMO) NPs, for which no similarity is detected with respect to any of the reference 2D templates (crystals or local symmetry). Visual inspection confirms that the systems characterized by low repulsion intensity ($\varepsilon_{pp}$ =

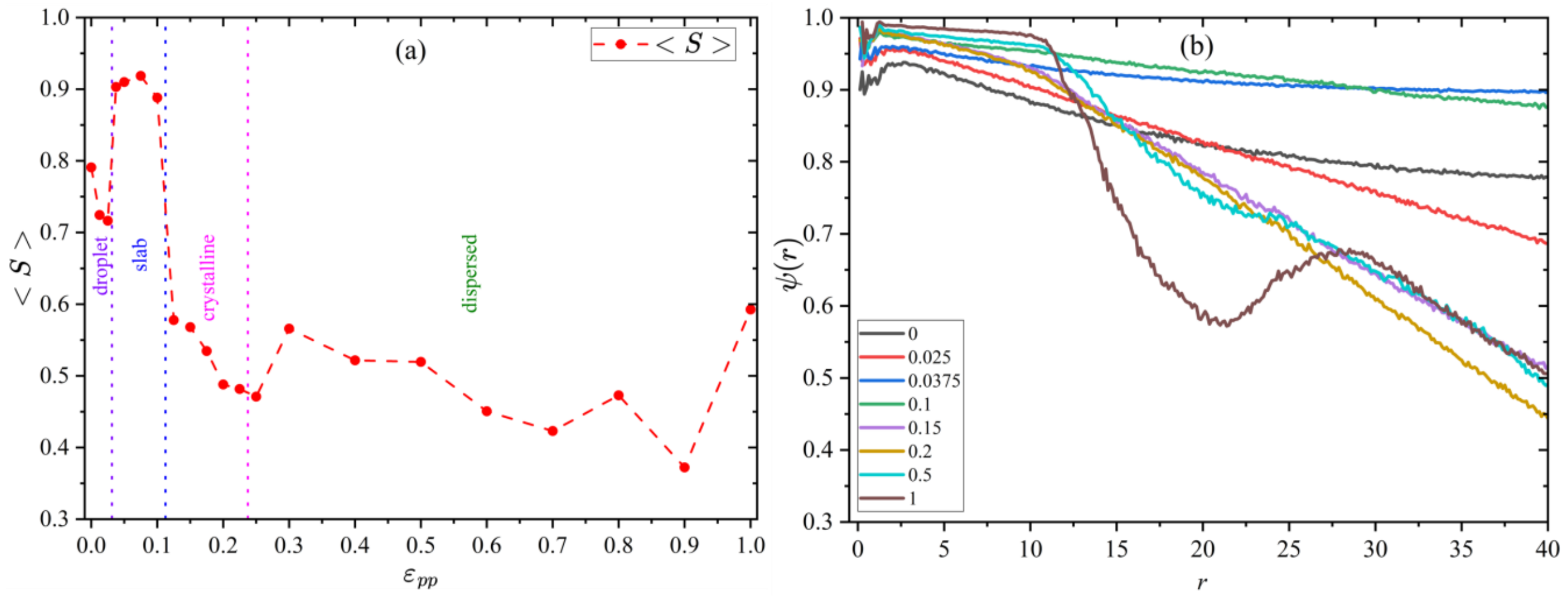

Figure 8. (a) Nematic order parameter, $\langle S\rangle$, of the polymer chains as a function of repulsion strength, $\varepsilon_{pp}$. Vertical dotted lines identify the crossovers between different NP morphologies denoted as "droplet", "slab", "crystalline droplet" and "dispersed". Dashed lines connecting the data points serve as guide for the eye. (b) Radial distribution of nematic order parameter, $\psi(r)$ as a function of inter-polymer distance, $r$ for selected $\varepsilon_{pp}$ values.

0.0125 and 0.05) in the droplet and slab phases show no traces of local ordering. In contrast, for the system at $\varepsilon_{pp}$ = 0.15 (crystalline droplet) most of the NPs that lie in the interior part of the formed cluster show a similarity that corresponds to the triangular crystal. Crystallization drops with increasing repulsion between NPs as mixing is favored. Accordingly, the population of interfacial NPs increases significantly while the fragmented clusters are limited in size and only partially crystallize. The dominant type of crystal for the dispersed phase is again the triangular one, which further corresponds to the densest monolayer packing made of hard spheres.

### *3.4. Self-Organization: Nematic Ordering of Polymers*

In the previous sections, we focused on the NP phase behavior as a function of inter NP repulsive strength $\varepsilon_{pp}$. However, the evolution of NP morphology strongly affects the overall nematic ordering and the spatial nematic correlation. Our past work on a similar system, as reported by Roy *et* al.[57], showed that strongly repulsive NPs ($\varepsilon_{pp}$ = 1), termed as "Hard NPs", disrupt the nematic ordering, whereas the non-interacting NPs ($\varepsilon_{pp}$= 0) termed as "Ideal NPs", facilitate nematic ordering. Here, we systematically explore the effect of intermediate repulsive interaction $0 \le \varepsilon_{pp} \le 1$ on nematic behavior. Towards this, the degree of nematic alignment of the polymer chains is quantified by the average nematic order parameter $\langle S\rangle$. The spatial extent of the ordering is characterized by the orientational correlation function, $\psi(r_{kl})$, where $r$ is the distance between the centers of mass of polymer chains $k$ and $l$. The orientational alignment of polymer end-to-end chains within the system is measured by $\langle S\rangle$. In order to calculate $\langle S\rangle$, first, end-to-end bond vector is defined for each polymer chain. Then, the angular difference $\Delta\theta_{kl}$ is computed to evaluate the pairwise nematic order as $S_{kl} = 2cos^2(\Delta\theta_{kl}) - 1$ for every possible pair in each configuration. The average nematic order parameter $\langle S\rangle$ is then sampled over all pairs and configurations:

$$\langle S\rangle = \frac{2}{n_{frames}M(M-1)}\sum_{f=1}^{n_{frames}}\sum_{k=1}^{M-1}\sum_{l=k+1}^{M} S_{kl}(f) \qquad (11)$$

where $n_{frames}$ is the number of frames and $M$ is the number of polymer chains in a configuration. The orientational correlation function $\psi(r_{kl})$, is calculated from the average of $2cos^2(\Delta\theta_{kl}) - 1$, considering all polymer pairs whose separation lies between $r$ and $r + \delta r$ over each configuration, following the process described in [57]. In the above calculations we set $\delta r = 0.025$.

The dependence of the nematic order parameter, $< S >$, on repulsion strength, $\varepsilon_{pp}$, is presented in Figure 8(a). Our results for the Ideal NPs ($\varepsilon_{pp}$= 0) confirm past observations[83] that in this limit nematic ordering is favored as indicated by $< S > \approx 0.80$. As repulsion increases in the droplet regime, a drop is observed in $< S >$. This drop is related to the increase in the droplet size which causes the loss of long-range ordering as also evidenced by the faster decay of $\psi(r)$ as shown in Figure 8(b) for $\varepsilon_{pp} = 0 \; and \; 0.025$. However, a larger droplet size also leads to a compact packing for the polymer segments, as verified by the Voronoi volume evolution in Figure 4(a), enhancing local nematic order[83]. Effectively, there is a trade-off between denser monomer packing, which enforces nematic ordering and a longer NP-monomer interfacial contour which interrupts it.

As the NP assembly turns to a slab morphology, the nematic order parameter reaches its maximum, with $< S > \approx 0.90$.This increment is attributed to two factors. First, stronger NP repulsion expands NP accessible volume, as evidenced by the sharp increase in $V_P$ and concomitant decrease in $V_m$. Thereby by confining the polymers into more compact domains, short-range nematic correlation is enhanced. Second, the change of NP cluster geometry from a curved droplet into a rectangular slab, whose almost straight contour allows the alignment of the neighboring polymer chains, supporting the long-range nematic correlation. The enhancement of the nematic correlation across the entire separation range is clearly seen by the higher values of $\psi(r)$ for $\varepsilon_{pp} = 0.1$ (Figure 8(b)).

As the NP repulsion increases further, a sharp drop in $< S >$ is triggered reducing to a value ~ $0.55$. This drop coincides with the transition of the NP assembly from slab to a rhombus-shaped crystalline droplet, whose sharp bends between the edges disrupt orientational coherence among the polymers, leading to a significant loss in long-range nematic correlation, as can be seen from the sharp decay in $\psi(r)$ for $\varepsilon_{pp} = 0.15 \; and$ 0.2 (Figure 8(b)). At even higher repulsion intensities, the NP droplet becomes progressively fragmented into smaller domains, while $< S >$ remains stabilized around $\sim$ 0.5, showing appreciable fluctuations as also noticed in other structural metrics. Notably, in this range, smectic polymeric domains emerge which gradually grow in extent with $\varepsilon_{pp}$, as evident from the corresponding snapshots in Figure 1. The smectic domains are locally strongly nematic, but their random orientations lead to loss in long-range nematic correlation. Consequently, $\psi(r)$ exhibits periodic undulations and long-range decay, as clearly shown for $\varepsilon_{pp}$ = 0.5 and 1.

## 4. CONCLUSIONS

Through Langevin Dynamics simulations we systematically analyze the effect of inter-nanoparticle intensity on the self-organization in polymer nanocomposites under extreme confinement. Polymers are represented as linear, semiflexible, rod-like chains of monomers of uniform size ($\sigma_m$), while nanofillers are spheres whose diameter is significantly larger than the one of monomers ($\sigma_p = 2.5\ \sigma_m$). Confinement is realized through the presence of flat and impenetrable walls along one dimension. The inter-wall distance is commensurate with the nanoparticle size ($L_z$ = 1.6 $\sigma_p$ = 4 $\sigma_m$), corresponding to extreme confinement that essentially allows the formation of a nanoparticle monolayer. Particles interact through the purely repulsive Weeks-Chandler-Andersen (WCA) potential, where the intensity of the monomer-nanoparticle and monomer-monomer interactions is kept fixed while the nanoparticle-nanoparticle interaction strength, $\varepsilon_{pp}$ is varied.

In all simulations, the initial configuration corresponds to a compact droplet-like clustering surrounded by polymer chains. As inter-nanoparticle repulsion increases, self-organization leads to either phase separation in polymer-rich and nanoparticle-rich domains or mixing. In more detail, we identify the presence of four distinct regimes with respect to the established self-assemblies: “droplet”, “slab”, “crystalline droplet” and “dispersed”. Cluster-based analysis reveals that in the first two regimes the initial nanoparticle droplet retains its structural integrity while fragmentation is established in the late part of the crystalline droplet regime and is completed in the dispersed regime.

The local crystalline order of the nanoparticles and liquid crystalline ordering of the polymers, are quantified through the Characteristic Crystallographic Elements (CCE) norm[73,74], and the nematic order parameter[55,57], respectively. Application of these refined metrics allows for a detailed analysis of the structural characteristics in each region. The "slab" phase is characterized by negligible crystallization, attaining its minimum value ($\tau^c \rightarrow 0$), as the nanoparticle cluster remains amorphous, while the polymers exhibit the highest nematic ordering $< S >\approx 0.90$. In contrast, the "crystalline droplet" regime is characterized by high local nanoparticle order, attaining maximum crystallinity ($\tau^c \rightarrow 0.55$), with extensive close-packed crystallites of triangular character, accompanied by a sharp reduction of nearly 40% to $< S >\approx 0.55$ in the nematic order parameter. In the "crystalline droplet" regime, the high triangular crystalline order results in a rhomboid-shaped NP assembly. Consequently, polymer chains are constrained to align along the sharp edges of the droplet, leading to a reduction in nematic ordering. Strong crystallization corresponds to an increase in the local volume around the nanoparticles, while the corresponding volume distribution becomes very narrow. The final "dispersed" phase is characterized by the fragmentation of the original droplet into small clusters, facilitating efficient mixing that leads to dense polymer packets and induces local smectic micro-domains.

This work demonstrates, at a fundamental level, that precise tuning of the interactions between the constituent components can successfully control the self-assembled morphologies in polymer-based nanocomposites under extreme confinement. Future efforts will focus on investigating the effect of varying nanoparticle concentrations and the degree of spatial confinement.

## 5. SUPPLEMENTARY MATERIAL

See the supplementary material for the density profiles of NPs and monomers along the confinement direction (S1), the NP-monomer pair distribution function (S2), the normalized CCE-based crystallinity (S3) and the evolution of CCE order parameters versus timesteps (S4).

## 6. AUTHOR CONTRIBUTIONS

P.C.: Data curation, Formal Analysis, Investigation, Methodology, Software, Validation, Visualization, Writing-original draft preparation. S.R.: Investigation, Conceptualization, Methodology, Formal analysis, Resources, Funding, Writing-review and editing, Supervision. S.D.: Investigation, Writing- review and editing. N.C.K.: Methodology, Formal analysis, Data curation, Funding acquisition, Resources, Writing-review and editing. All authors have read and agreed to the published version of the manuscript.

## 7. ACKNOWLEDGEMENTS

The support through projects "PID2021-127533NB-I00" of MICINN/FEDER (Ministerio de Ciencia e Innovación, Fondo Europeo de Desarrollo Regional) and "Seeds Project-VRISEM25NK", "Seeds Project-VRISEM26NK" and "D05DMF26" of UPM (Universidad Politécnica de Madrid) is deeply appreciated. N.C.K. gratefully acknowledges the Universidad Politécnica de Madrid (www.upm.es) for providing computing resources on the Magerit Supercomputer. The fruitful discussions with Daniel Martínez-Fernández, Katerina Foteinopoulou, Zhou Wan and Fernando Monson are deeply appreciated. S.R. gratefully acknowledges Prof. Yeng Long Chen, Institute of Physics, Academia Sinica (www.phys.sinica.edu.tw), and Dmytro A. Luzhbin, Institute of Statistical Science, Academia Sinica (www.stat.sinica.edu.tw) for their valuable discussions and support.

## 8. CONFLICT OF INTEREST

The authors declare no conflicts of interest.

## 9. DATA AVAILABILITY

The data that support the findings of this study are available from the corresponding author upon reasonable request.

# Supplementary Material

# Inter-Nanoparticle Repulsion Effect on Self-Organization in Polymer Nanocomposites: A Langevin Dynamics Simulation Study

Paawan Chandrakanth,[1] Nikos Ch. Karayiannis,[2] Subhalaxmi Das[1] and Supriya Roy[1,a]

[1]School of Applied Sciences, Kalinga Institute of Industrial Technology, Bhubaneswar, Odisha, India

[2] Institute for Optoelectronic Systems and Microtechnology (ISOM) and Escuela Técnica Superior de Ingenieros Industriales (ETSII), Universidad Politécnica de Madrid (UPM), José Gutiérrez Abascal 2, E-28006 Madrid, Spain

[a] Author to whom correspondence should be addressed: supriyophy163@gmail.com

## S1. Density Profiles

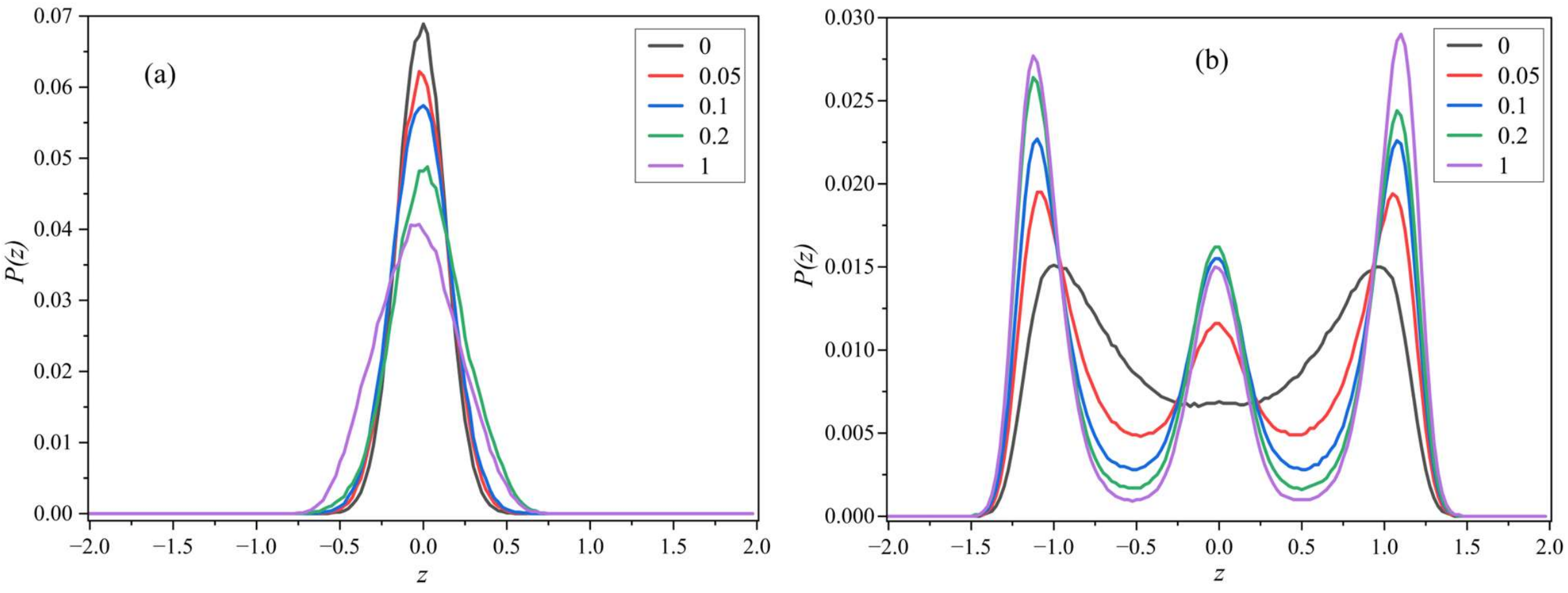

Figure S1: Density profiles along the confined $z$-dimension of (a) nanoparticles and (b) monomers for different repulsion strength, $\varepsilon_{pp}$ as indicated in the legend.

Figure S1(a) shows the density profiles of NPs along the confined $z$ dimension for different $\varepsilon_{pp}$ values. A bin width of $\delta z$ = 0.025 is implemented. $z$ = 0 corresponds to mid-plane while flat impenetrable walls exist at $z$ = ±2. In all systems, nanoparticles lie in the interval $-0.75 \leq z \leq +0.75$, i.e. the NP centers remain at a distance of at least $\sigma_p/2$ from each wall lying practically in the accessible central stripe in the $z$ dimension, of width $L_z - \sigma_p = 1.5$. Therefore, this justifies the need for the calculation of the effective volume fraction of nanoparticles (Eq. (8) in main text). Based on the above, the NPs form a single, quasi-2D layer symmetrically centered around the middle of the slit ($z = 0$) for all repulsion strengths. For $\varepsilon_{pp} = 0$, the density profile is the sharpest and as $\varepsilon_{pp}$ increases the peak intensity systematically decreases, while the profile broadens. This indicates fluctuations of the NPs around the mid-plane as the NP assembly expands and eventually disperses. As the single-layer arrangement is preserved over the whole range of $\varepsilon_{pp}$ it justifies our usage of the *x*-*y* projections in the 2-D pair distribution functions, $g_{pp}(r)$ in the main text, and in the CCE analysis of the main text.

Figure S1(b) presents the corresponding $z$-density profile for monomers. Accordingly, monomers lie in the interval density of $-1.5 \leq z \leq +1.5$, which corresponds to an accessible central stripe with width $L_z - \sigma_m = 3$. For $\varepsilon_{pp} = 0$, the profile displays two symmetric broad maxima at $z \approx \pm 1.0$, with a shallow minimum at the center. Upon activating repulsion between NPs, a distinct layered arrangement emerges with two peaks adjacent to each wall at approximately $z \approx \pm 1.1$ and a third one at the mid-plane $z$ = 0. As repulsion strength increases, the peaks shift slightly closer to the walls and become more pronounced while the observed minima located at $z \approx \pm 0.5$ become deeper. This trend points towards partitioning of the system into three distinct layers, driven by a reduction in available volume for monomers due to NP dispersion.

Combining the information of the mass density profiles along the *z* dimension we can further conclude that the confining agents act as effectively impenetrable walls, as intended.

### S2. Radial Distribution Function for NP-monomer pairs

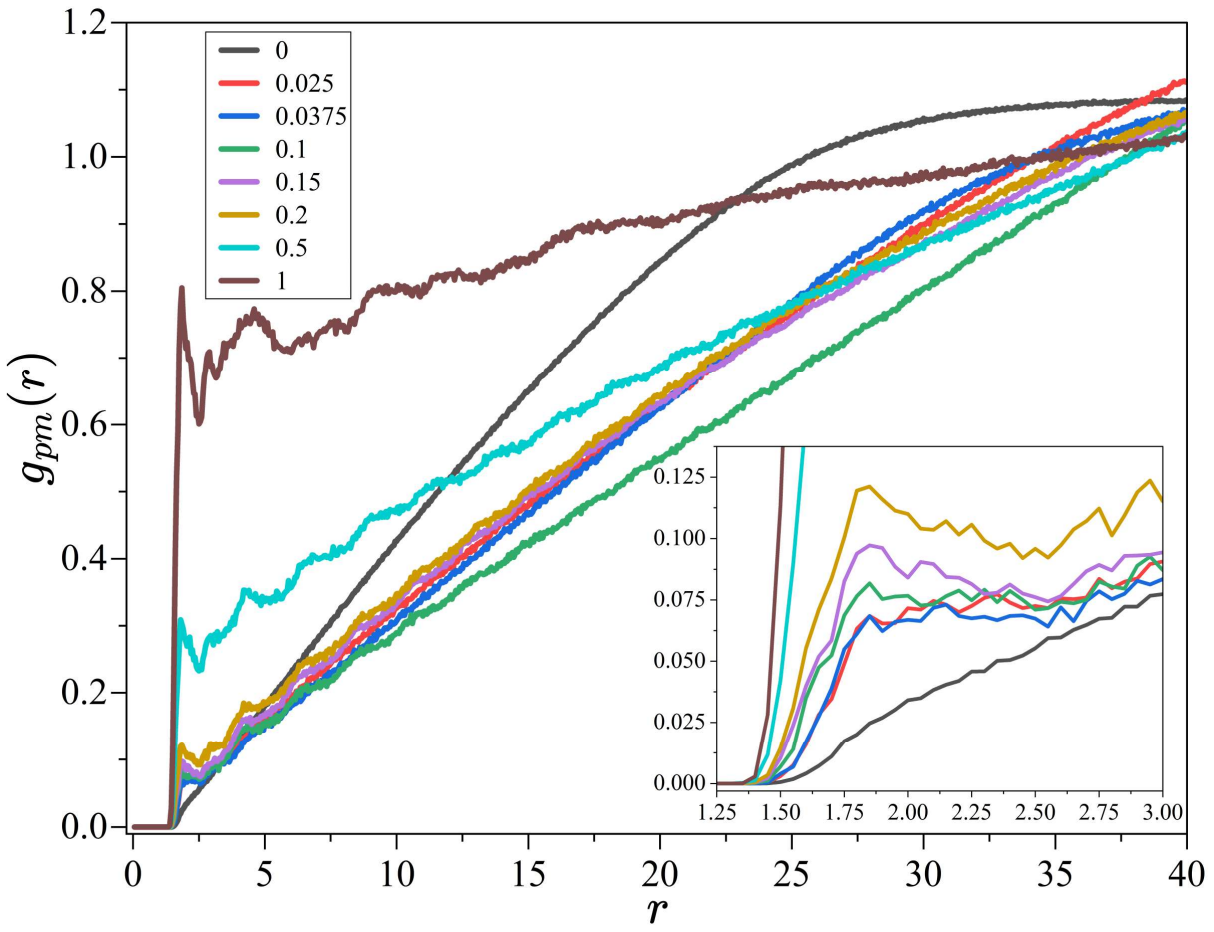

Figure S2. Pair radial distribution function (RDF) between nanoparticles and monomers, $g_{pm}(r)$, as a function of distance, $r$, for different repulsion strengths, $\varepsilon_{pp}$, as indicated in the legend. Inset: magnification of the short-range region ($1.25 \leq r \leq 3.0$).

Figure S2 displays the pair radial distribution function (RDF) between NPs and monomers $g_{pm}(r)$, for different $\varepsilon_{pp}$. For all systems, $g_{pm}(r)$ vanishes at short distances $(r \leq 1.3)$, due to the excluded-volume interactions between NPs and monomers. As shown in the inset, $g_{pm}(r)$ becomes non-zero at around $r \approx 1.4 - 1.5$, slightly below the tangency condition for NP-monomer pairs, $r_{contact} = \frac{(\sigma_p + \sigma_m)}{2} = 1.75$, reflecting the soft character of the WCA potential.

For $\varepsilon_{pp} = 0$, no contact peak is observed and $g_{pm}(r)$ increases smoothly and levels at a stable value at $r \approx 30$. The plateau at $r \approx 30$ is consistent with the size of the compact NP droplet observed, as also reflected by $g_{pp}(r)$ in Fig. 6(a) of the main text. In the droplet, slab and crystalline droplet regimes $(0.025 \leq \varepsilon_{pp} \leq 0.2)$, a weak first peak appears at $r \approx 1.8$, slightly above $r_{contact}$. With increasing repulsion strength, the peak height increases systematically, as seen in the inset. Initially, this peak can be attributed from nanoparticle-monomer pairs at the interface of the droplet. As the self-assembled cluster begins to expand the population of such pairs increases as does the contour of the interface. In the dispersed regime, corresponding to the systems of $\varepsilon_{pp}$ = 0.5 and 1 in Fig. S2, the contact peak increases to much higher values, as the NP dispersion and the corresponding intermixing lead to abundance of dissimilar pairs near contact.

## S3. Normalized CCE-based Crystallinity

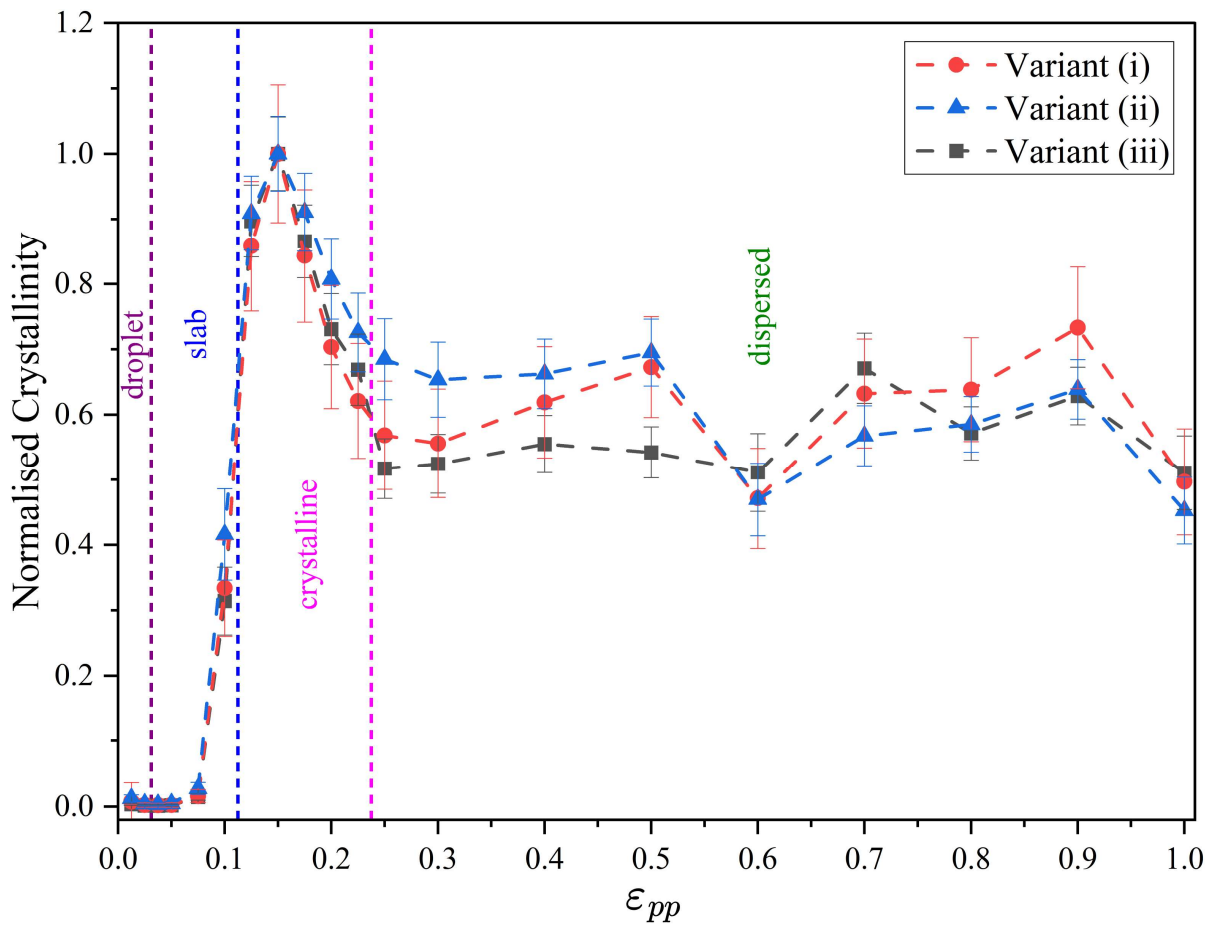

Figure S3. Degree of crystallinity, $\tau^c$, normalised by its value at $\varepsilon_{pp} = 0.15$, as a function of repulsion strength, $\varepsilon_{pp}$, for the three variants of the CCE analysis: (i) $\varepsilon^{\{thres\}} = 0.245$ (red circles), (ii) $\varepsilon^{\{thres\}} = 0.305$ (blue triangles) and (iii) $\varepsilon^{\{thres\}} = 0.305$ with the normalization factor $n^{CCE}$, calculated with $r_{min} = 3.75$ (black squares). In all cases $l_{CCE} = \sigma_p = 2.5$. For the definitions see corresponding description in the main text. Vertical dotted lines identify the crossovers between the different NP morphologies.

To analyze the sensitivity of the crystallographic analysis (CCE) to parameter choice, the normalized degree of crystallinity is calculated for three variants of the CCE norm: variant (i): $\varepsilon^{\{thres\}} = 0.245$; variant (ii): $\varepsilon^{\{thres\}} = 0.305$ and variant (iii): $\varepsilon^{\{thres\}} = 0.305$ with the crystallinity multiplied by the normalization factor $n^{CCE}$ defined in Eq. (10) of the main text. In all cases it stands $l_{CCE} = \sigma_p = 2.5$. For variant (iii), the integral of $g_{pp}(r)$ in $n^{CCE}$ extends to its first minimum $r_{min} = 3.75$, i.e., over the first coordination shell of the NPs. In absolute terms, the stricter threshold of variant (i) yields a systematically lower crystallinity with a maximum $\tau^c \approx 0.26$ compared to $\tau^c \approx 0.53$ for variant (ii). To compare these trends directly, the crystallinity for each variant is normalised by its maximum value, which in all cases is encountered at $\varepsilon_{pp} = 0.15$ in Fig. S3.

All three CCE norm variants yield the same phase behavior of the NPs as a function of repulsion strength. In the droplet and slab regimes $(0.0125 \leq \varepsilon_{pp} \leq 0.05)$ the NP assemblies are amorphous, with normalized crystallinity below 0.015 in all cases. The first crystal nuclei appear at $\varepsilon_{pp} = 0.075$, and a sharp increase in local order takes place at $\varepsilon_{pp} = 0.1$, at the transition from the slab to the crystalline droplet. All three variants exhibit their maximum crystallinity at $\varepsilon_{pp} = 0.15$, followed by a systematic decrease within the crystalline droplet regime. In the dispersed regime, the normalized crystallinity remains systematically below the maximum. The differences between the three CCE variants exceed the statistical uncertainty at some repulsion strengths, reflecting the larger fluctuations of the cluster morphology in this regime. The normalization factor $n^{CCE}$ leaves the trends in the droplet, slab and crystalline droplet regimes practically unchanged, and modifies only the values in the dispersed regime, where $n^{CCE}$ itself varies significantly (≈ 8.6 – 13.5) with $\varepsilon_{pp}$.

Based on the above, the identification of the four regimes and of the maximum crystallinity at $\varepsilon_{pp} = 0.15$ is independent of the choice of CCE norm parameters. The same holds for all qualitative trends with respect to crystallization. Accordingly, the analysis in the main text is based on variant (ii).

## S4. CCE order parameters versus time steps

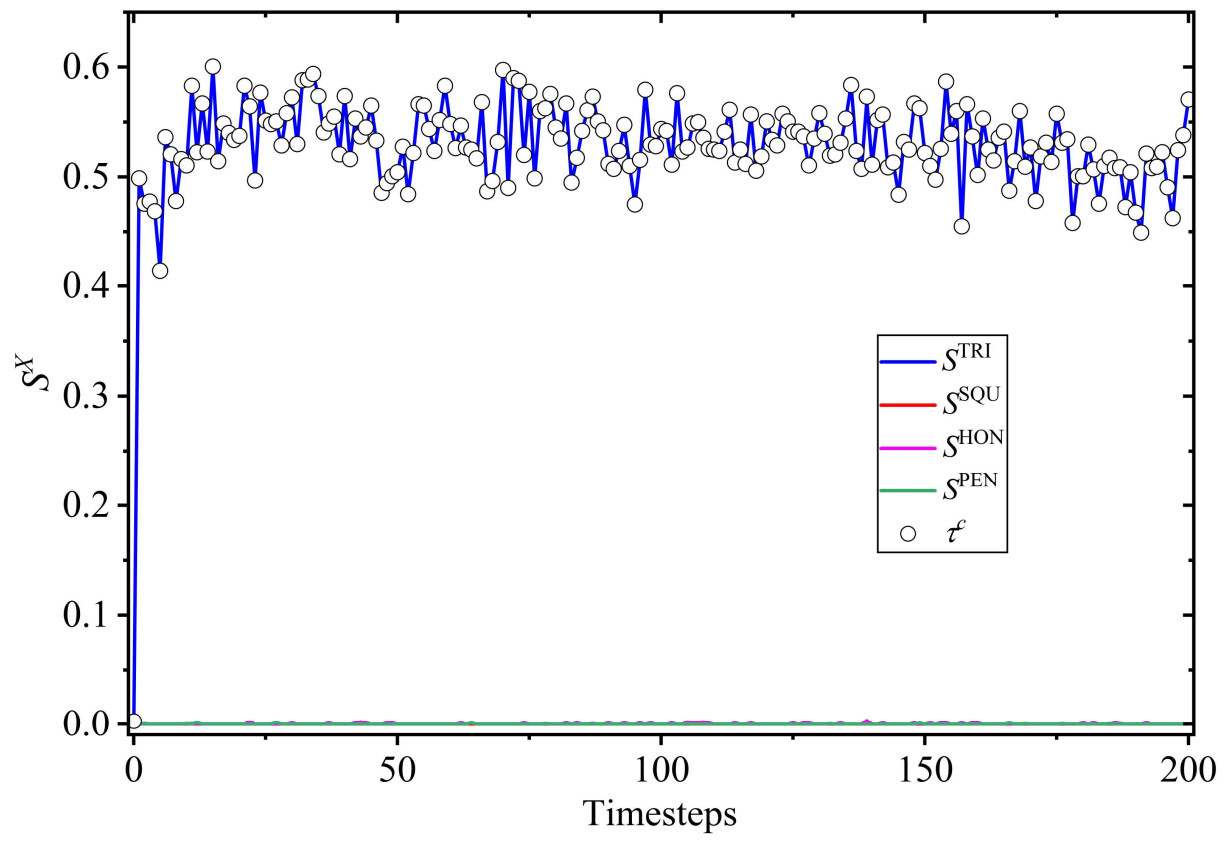

Figure S4. Evolution of the CCE order parameters, $S^X$ ($X$ = TRI, SQU, HON and PEN), and of the total degree of crystallinity, $\tau^c$, as a function of time steps for the system with $\varepsilon_{pp} = 0.15$. Consecutive configurations are separated by 2 × $10^6$ time steps. The order parameters have been calculated employing $\varepsilon^{thres} = 0.305$ and $l_{CCE} = 2.5$.

Figure S4 shows the evolution of the CCE order parameters, $S^X$, and of the total degree of crystallinity, $\tau^c$, for the system with $\varepsilon_{pp} = 0.15$, which exhibits the highest crystallinity among all systems studied here. Once the repulsion between NPs is activated, crystallinity rises sharply, reaching $\tau^c \approx 0.5$ already in the first frame (2 × $10^6$ time steps). Then, it fluctuates within $0.45 \leq \tau^c \leq 0.6$ for the rest of the trajectory. As can be seen by the evolution of the CCE order parameters, the ordered morphology corresponds exclusively to triangular (TRI) crystals, since $S^{SQU}, S^{HON}$, and $S^{PEN}$ remain practically zero throughout the simulation. Thus, $\tau^c$ coincides with $S^{TRI}$.